\pdfoutput=1
\documentclass[%
 reprint,
 amsmath,amssymb,
 aps,
 prper,
]{revtex4-1}

\usepackage{graphicx}
\usepackage{dcolumn}
\usepackage{bm}
\usepackage{color,soul}
\usepackage{float} 
\usepackage{subcaption,natbib,multirow}

\begin{document}

\preprint{APS/123-QED}

\title{Characterizing active learning environments in physics using latent profile analysis}

\author{Kelley Commeford}
\author{Eric Brewe}%
\affiliation{%
 Drexel University\\
 3141 Chestnut St, Philadelphia, PA 19104%
}

\author{Adrienne Traxler}
\affiliation{
 Wright State University\\
 3640 Colonel Glenn Hwy, Dayton, OH 45435
}%

\date{\today}

\begin{abstract}
The vast majority of research involving active learning pedagogies uses passive lecture methods as a baseline. We propose to move beyond such comparisons to understand the mechanisms that make different active learning styles unique. Here, we use COPUS observations to record student and instructor activities in six known styles of active learning in physics, and use latent profile analysis to classify these observations. Latent profile analysis using two profiles successfully groups COPUS profiles into interactive lecture-like and other. Five latent profiles successfully sorts observations into interactive lecture-like, Modeling Instruction, ISLE labs, Context-Rich problems labs, and recitation/discussion-like. This analysis serves as a proof of concept, and suggests instructional differences across pedagogies that can be further investigated using this method. 
 
\end{abstract}

\pacs{Valid PACS appear here}
\maketitle

\section{Introduction}
Active learning has been gaining traction in physics classrooms, with numerous studies showing how active learning can promote learning gains over passive lecture alternatives. However, within the umbrella of `active learning', several physics pedagogies have spawned from different ideological foundations, often with wildly varying mechanical differences \cite{Meltzer2012ResourcePhysics,redish_research-based_2007,thacker_recent_2003}. As such, the term `active learning' no longer sufficiently describes a pedagogy, making it difficult to pinpoint the mechanisms that lead to observed learning gains.  Thus, it is imperative that we understand the mechanisms that make different `active learning in physics' pedagogies unique, so future studies can further delineate the benefits of specific active learning mechanisms. 


The `Characterizing active learning environments in physics' project, or CALEP, sets out to establish a vocabulary that will allow us to speak about the different active learning pedagogies in physics, without relying on a comparison to passive lecture methods. We focused on two aspects of the active learning pedagogies, how students and instructors spend time in class, and the student networks that result from the instruction. A full summary of the CALEP project can be found in \cite{Commeford2020CharacterizingObservations}. We collected classroom observations and student network data from six prominent research-based introductory physics curricula. For the classroom observations, we used the Classroom Observation Protocol for Undergraduate STEM (COPUS)\cite{Smith2013}, which allows us to measure the time dedicated to certain activities during a class-period, which, in essence, gives us an idea of the kinds of activities that occur in different active learning environments. For network analysis results, please refer to \cite{Commeford2020CharacterizingObservations, Traxler2020NetworkPhysics}.

This paper will focus on the COPUS observations, where we discuss the results of latent profile analysis (LPA) on the observed COPUS profiles. Latent profile analysis can be used to tease out hidden categories within your data-- such as physics pedagogies. LPA has been used with COPUS data to describe a large sample of STEM classes, which effectively sorted them into active, passive, and hybrid categories~\cite{Stains2018AnatomyUniversities}. Our analysis builds on that work by restricting to known active learning pedagogies in physics classrooms, as a first effort at developing a more nuanced classification scheme.

\section{Background}

Because active learning is an umbrella term for different pedagogical approaches, it fails to identify distinguishing characteristics of pedagogies in physics. To this end, we identified six pedagogies that are well represented in the literature and at professional development workshops at national conferences. These six pedagogies have different approaches to active learning. In order to explore these different approaches, we used COPUS as a means to understanding unique features of classroom activity. 

\subsection{Active learning pedagogies in physics}

We collected data from high fidelity implementations of six well known active learning pedagogies in physics. The pedagogies we studied were Tutorials in Introductory Physics, the Investigative Science Learning Environment (ISLE), Modeling Instruction, Peer Instruction, Context Rich Problems (aka the Minnesota curriculum), and SCALE-UP.

Tutorials in Introductory Physics maintains a lecture/lab/recitation structure typical of large introductory courses. The recitations, or tutorial sections, are interactive small group environments that have the students follow scaffolded worksheets that emphasize misconception confrontation and resolution. The teaching assistants facilitate small group interactions~\cite{McDermott2002,Scherr2007EnablingMaterials}. 

ISLE (Investigative Science Learning Environment) treats students as neophyte scientists, and focuses on building up correct intuition rather than debunking misconceptions~\cite{Etkina2007InvestigativePhysics,Etkina2015MillikanPractices}. Students work in small groups through a lab-based learning cycle, while the instructors facilitate discussion. 

Modeling Instruction is a fully contained learning environment where students work in small groups to develop, test, deploy, and revise conceptual models, and then come together for large `white-board meetings' to discuss their findings~\cite{PhysPortInstruction,Wells1995AInstruction,Brewe2008}. The instructors facilitate small group discussion and guide the larger white-board meetings.

Peer Instruction is a structured, interactive lecture environment, typically used when classroom constraints deem a large lecture hall necessary. Peer Instruction divides a class session into small modules, which begin with a short lecture, and cycle through several clicker-questions. Students answer individually and then re-answer after discussing with their peers~\cite{Mazur1997,Crouch2007PeerOnce}. 

Context-rich problems leads students through the course under a pseudo apprentice-expert relationship. This pedagogy uses context-rich problems in all aspects of the course, to teach students how to identify relevant information and engage in expert-like problem solving behavior, as demonstrated by the instructor and teaching assistants~\cite{UniversityDevelopment,Heller1992TeachingGroups}. 

Finally, SCALE-UP is also a fully contained classroom environment, where the students are in pairs or groups of three, but seated at large tables with other groups of two or three students to facilitate larger discussions. SCALE-UP usually refers more to the classroom setup than the pedagogy itself, but lends itself nicely to the adoption of many active learning styles~\cite{Beichner2007TheProject}. As such, it is often treated as an independent pedagogy. 

\subsection{The Classroom Observation Protocol for Undergraduate STEM}\label{sec:COPUS}
The Classroom Observation Protocol for Undergraduate STEM (COPUS) was developed by Smith et.al~\cite{Smith2013}. It consists of 12 instructor and 13 student codes, for which the observer marks whether the coded behavior occurred or not during two-minute intervals. An activity is counted if the behavior occurs for at least five seconds during the two-minute interval. A full list of COPUS code abbreviations and their descriptions has been included in the supplementary material. An example of a COPUS observation for a tutorial session can be seen in table \ref{tab:COPUSExample}.

\begin{table*}[th]
\caption{COPUS observation example. Time is measured in two-minute intervals.}
\label{tab:COPUSExample}
\begin{tabular}{c|ccccccccccccc|cccccccccccc}
 & \multicolumn{13}{c}{\textbf{Students Doing}} 
 & \multicolumn{12}{c}{\textbf{Instructor Doing}} \\
 \cline{2-26}
 \textbf{Time}& L & IND & CG & WG & OG & AnQ & SQ & WC & Prd & SP & T/Q & W & O & Lec & RtW & Fup & PQ & CQ & AnQ & MG & 1o1 & D/V & Adm & W & O \\
 \hline
 0-2   &&&& X &&&   &&&&&& X &&&&&&& X & X &&& X\\  \hline
 2-4   &&&& X &&& X &&&&&& X &&&&&&& X & X      \\  \hline
 4-6   &&&& X &&& X &&&&&& X &&&&&&& X & X      \\  \hline
 6-8   &&&& X &&&   &&&&&& X &&&&&&& X & X      \\  \hline
 8-10  &&&& X &&&   &&&&&& X &&&&&&& X & X      \\  \hline
 10-12 &&&& X &&& X &&&&&&   &&&&&&& X & X      \\  \hline
 12-14 &&&& X &&&   &&&&&&   &&&&&&&   & X      \\  \hline
 14-16 &&&& X &&& X &&&&&&   &&&&&&& X & X      \\  \hline
 16-18 &&&& X &&&   &&&&&&   &&&&&&& X & X &&& X\\  \hline
 18-20 &&&& X &&&   &&&&&&   &&&&&&& X & X      \\
 \hline
 \textbf{Total} & 0 & 0 & 0 & 10 & 0 & 0 & 4 & 0 & 0 & 0 & 0 & 0 & 5 & 0 & 0 & 0 & 0 & 0 & 0 & 9 & 10 & 0 & 0 & 2 & 0
\end{tabular}
\end{table*}

From these observations, we can compile the selected codes into COPUS profiles. The method of COPUS profile creation is not always clearly reported in literature that uses COPUS, making it difficult to compare analyses. It is typical of such studies to report percentages for each code, without a description of how ratios were taken or visualizations to infer the same information. As such, we explain our process here. For the CALEP project, we used the `bar chart' method, in which COPUS profiles were created by summing the number of marks in each column (indicated in the `total' row in table \ref{tab:COPUSExample}), and dividing by the number of intervals over which the observation occurred (10 intervals, using the same example). This method leads to the percentage of class time that a code was present. An example of a COPUS profile from CALEP can be seen in table \ref{tab:COPUSProfileExample}. It should be noted, however, that we calculate our COPUS profiles differently than in Smith et.al. In Smith et.al., the number of marks is tallied for each column, and divided by the total number of marks across all columns. This method, which we dub `pie chart' method, is useful when investigating the prevalence of activities. We chose the former method of profile creation, as we felt it provided a more explicit picture of how class time was spent. If an instructor spends the majority of the time lecturing, but peppers in several other instructional methods, the latter method of profile creation can be misleading and under-report the frequency of a code. We wanted a profile that showed the fraction of class time engaged in a certain instructional method, not fraction of every code reported. 

\begin{table}[th]
\caption{COPUS profile example. The total number of occurrences for each code was tallied in each column, and divided by the number of intervals. For this example, there were 10 intervals.\label{tab:COPUSProfileExample}}
\begin{ruledtabular}
\begin{tabular}{cc|cc}
 \multicolumn{2}{c}{\textbf{Students Doing}} 
 & \multicolumn{2}{c}{\textbf{Instructor Doing}} \\
 \hline
 L   & 0   & Lec & 0  \\
 IND & 0   & RtW & 0  \\
 CG  & 0   & Fup & 0  \\
 WG  & 1.0 & PQ  & 0  \\
 OG  & 0   & CQ  & 0  \\
 AnQ & 0   & AnQ & 0  \\
 SQ  & 0.4 & MG  & 0.9\\
 WC  & 0   & 1o1 & 1.0\\
 Prd & 0   & D/V & 0  \\
 SP  & 0   & Adm & 0  \\
 T/Q & 0   & W   & 0.2\\
 W   & 0   & O   & 0  \\
 O   & 0.5

\end{tabular}
\end{ruledtabular}
\end{table}

\subsection{Latent Profile Analysis}
Latent profile analysis (LPA) uses Gaussian mixture modeling to uncover hidden groupings, also known as classes or profiles, of data based on some measured variable~\cite{Oberski2016MixtureAnalysis}. For the CALEP project, our goal is to determine what the quantifiable identifying features of different physics pedagogies are. Active learning comes with an abundance of student-centered activities, which, in theory, are different for each pedagogy. As such, we chose to use LPA to see if this analysis method could identify different pedagogies based on observations of active learning behaviors. We used the percentage of class time per COPUS code (known as the COPUS profile) as our input variables to LPA, with the expectation that LPA would in turn group observations with similar COPUS profiles. Ideally, these groupings would align with the identified active learning pedagogies each observation represents.

\subsection{Previous studies using LPA}\label{sec:previousstudies}
Stains et. al. has previously utilized latent profile analysis with COPUS observations from several science disciplines~\cite{Stains2018AnatomyUniversities}. They collected COPUS observations from classes across STEM disciplines, most of which were at doctoral-granting institutions, using a convenience approach-- a call was put out on Discipline-Based Education Research (DBER) listservs asking researchers to share COPUS data, and they also collected voluntary COPUS submissions using the COPUS analyzer website~\cite{StainsCOPUSAnalyzer}. After using LPA on these solicited COPUS profiles, Stains ultimately chose the model with seven profiles for their data, which they further sorted into didactic (mostly lecture), interactive lecture (hybrid), or student centered (active) learning environments. 

Stains et al took random observations and THEN classified them into active, passive, or hybrid pedagogies. We differ from Stains et al in that we used purposeful sampling, collecting data at sites chosen due to their high fidelity implementation of different active learning pedagogies. Further, we feel that it is no longer sufficient to be abstract in classifying pedagogies as `active' or `not active', so our purpose was to extend the work of Stains et al by zooming in on the active learning category, while limiting our scope to investigate differences among active learning specifically in physics.

\begin{figure}[h]
    \includegraphics[width=\linewidth]{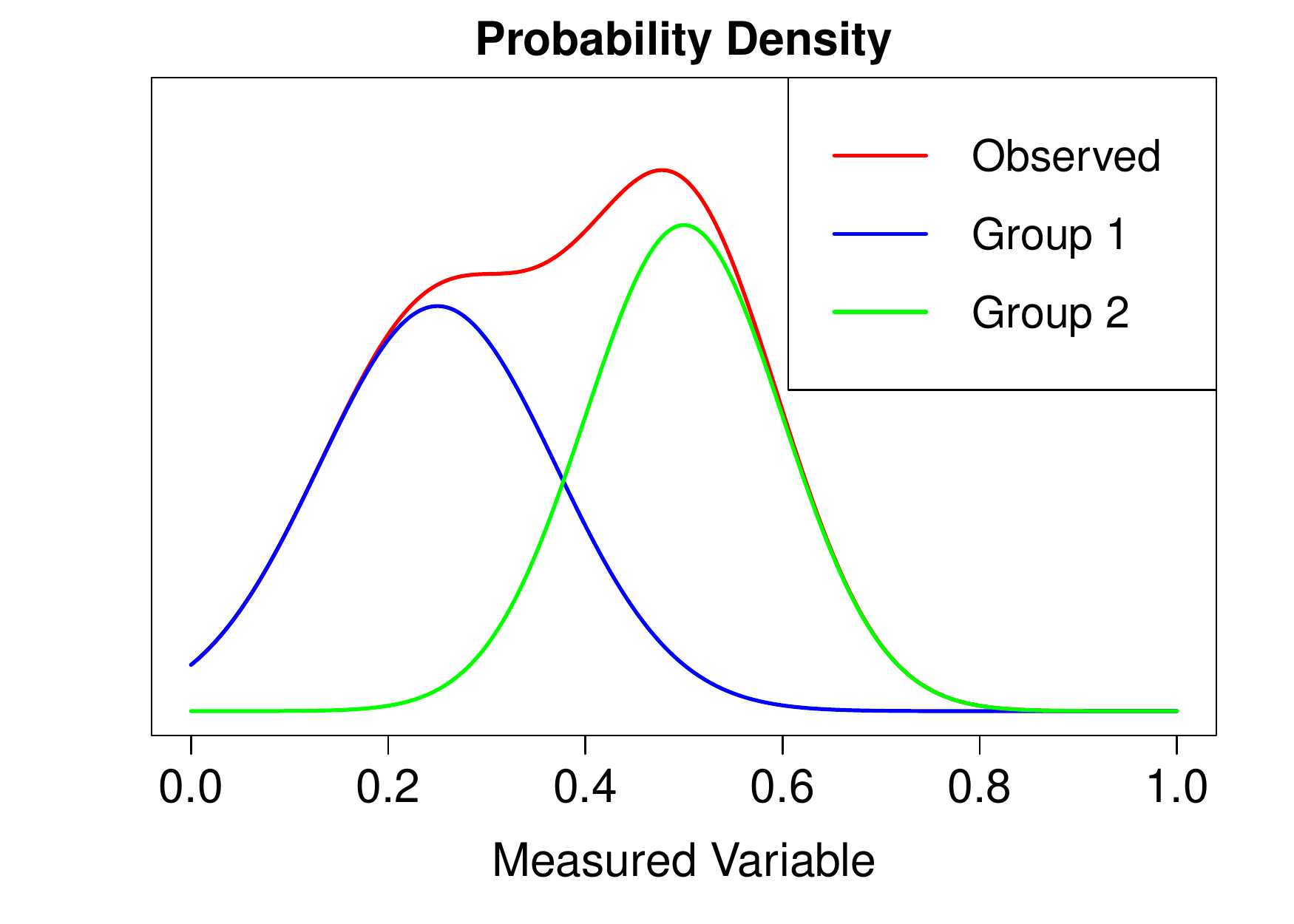}
    \caption{Given an observed distribution (red), LPA assumes that this observed distribution can be described by a combination of distinct profiles that follow a Gaussian distribution (two groups, shown in blue and green).}
    \label{fig:fakeLPA}
\end{figure}

We have provided a visual in figure \ref{fig:fakeLPA} to help illustrate how LPA classifies COPUS profiles into distinct profiles. Given a distribution of an observed variable (in our case a single category from COPUS), shown in red, LPA assumes that this observed distribution can be described by a combination of distinct profiles that follow a Gaussian distribution (two groups, shown in blue and green). LPA begins by taking some initial Gaussian distribution to represent each profile, of which you can have any number, then iterates through the data and adjusts the Gaussians until all points are assigned to a profile such that the expectation value is maximized. LPA does this by calculating a probability distribution for each profile (the group Gaussians), and assigns a weighted profile membership to each data point (via probability distribution). It then recalculates the probability distribution representing each profile using the weighted members, and repeats this process until it converges\footnote{Victor Lavrenko does in incredible job illustrating how this process works on his YouTube channel http://bit.ly/EM-alg }. 

LPA can also accept constraints to these Gaussians, via variance and covariance. The variance can be equal between profiles (the Gaussians have the same width), or vary between profiles. Additionally, the prevalence of input variables can depend on each other, introducing co-variance, which can be set to zero, equal, or varying between profiles in the model as well~\cite{Fraley2007BayesianClustering}. Understanding the variance of the profile distributions is a bit more straightforward if we preemptively reference the top panel of figure \ref{fig:FullList-Model1}; The variance is the size of the box, and equal variance forces the size of the boxes for each code to be equal for each profile. Varying variance, on the other hand, allows these boxes to be different sizes. The co-variance describes how the input variables depend on each other, which is a bit harder to visualize. In the context of our data, co-variance would tell us how the presence of lecture depends on the presence of real-time-writing. While co-variance is likely present in our data (by observation, real-time-writing does typically accompany lecture), we did not have a large enough data set to get the varying co-variance model to converge. Additionally, the equal co-variance model, which forces each pedagogy to follow the same relationships between codes, was excluded as it was not an appropriate constraint for our study. Thus, we assume zero co-variance in our analysis. Future work should explore the role that co-variance plays in the development of profiles, where a larger data set can illuminate meaningful relationships between COPUS codes.



\section{Methods}
We collected COPUS observations from six high fidelity active learning pedagogies in physics~\cite{Commeford2020CharacterizingObservations}. We traveled to the sites and observed as many sections of the course as possible in a one week visitation period. All observations were done in a live environment, in person, by the same observer. The official COPUS recording spreadsheet was used for these data collections~\cite{Smith2013}. A summary of the number of sections and the number of observations per section can be seen in table \ref{tab:numbersections}.

\begin{table}[h]
\caption{Number of course sections and observations for each pedagogy included in this study.}\label{tab:numbersections}
\begin{ruledtabular}
\begin{tabular}{ccc}
 \textbf{Pedagogy}& Num. Sections & Num. Obs.  \\
 \hline
  Tutorials & 1 Lecture & 1 \\
  & 19 Tutorials & 1\\
  \hline
  ISLE Whole Class  & 1 Lab & 1 \\
  & 4 Recitations & 1\\
  & 1 Lecture & 2\\
  \hline
  ISLE Lab Only & 5 Labs & 1\\
  \hline
  Modeling Instruction & 3 & 2 \\
  \hline
  Peer Instruction   & 1     & 1    \\
  \hline
  Context-Rich Problems & 2 Discussions & 1\\
  & 4 Labs & 1 \\
  & 2 Lectures & 2 \\
  \hline
  SCALE-UP & 1 & 3
\end{tabular}
\end{ruledtabular}
\end{table}

From the base COPUS observations with checkmarks in each code column, we created the COPUS profiles by summing the number of checks per column and then dividing by the number of observation intervals (see section \ref{sec:COPUS} for a more thorough explanation of profile creation).

For this analysis, individual class period observations are left as stand-alone COPUS profiles. We originally wanted to cluster observations into a single profile to represent a week's worth of class time for each pedagogy. This means, for example, an ISLE course that includes a lecture, recitation, and lab, would be reported as one COPUS profile that encapsulates the entire pedagogical experience for that week. Unfortunately, grouping observations in this way ensured we did not have enough data points for the LPA models to converge. As such, we left the COPUS profiles as individual class period observations, separated into various course components. 

\subsection{TidyLPA in R}

There are six different LPA Gaussian mixture models that can be used in the TidyLPA package in R~\cite{tidylpa}, two of which are run using wrappers to the proprietary software, Mplus~\cite{Muthen1998MplusGuide}. As such, those two models are not included in this analysis. 

Each model specifies how the variance and co-variance of the Gaussian mixture models is altered between iterations. For equal variance, the variance of the fit Gaussians for each profile are assumed to be equal across profiles. The co-variance for inter-COPUS code relationships can be set to zero, equal between profiles, or varying between profiles. Table \ref{tab:ModelVariance} illustrates the variance and co-variance combinations available, and their model number in TidyLPA.

\begin{table}[h]
\caption{Variance and co-variance of TidyLPA model. Models 4 and 5 were not included, as they are run using the proprietary software, Mplus~\cite{Muthen1998MplusGuide}\label{tab:ModelVariance}}
\begin{ruledtabular}
\begin{tabular}{c|ccc}
 & \multicolumn{3}{c}{\textbf{Co-variance}} \\
 \textbf{Variance}& Varying    & Equal    & Zero    \\
 \hline
  Equal   & MPLUS      & 3        & 1       \\
  Varying & 6          & MPLUS    & 2     \\
\end{tabular}
\end{ruledtabular}
\end{table}

In this paper, we show the results from model 1 analysis. Due to our relatively small data set, models 2 and 6 fail to converge. Model 3, which assumes equal co-variance between observed variables for each profile, was ultimately discarded as we felt it was not appropriate to force that condition. It would, however, be more interesting and realistic to use models 2 or 6, but we require more observations in order to get a convergent result.

\section{Results}
\begin{figure*}[!]
    \includegraphics[height=0.4\textheight]{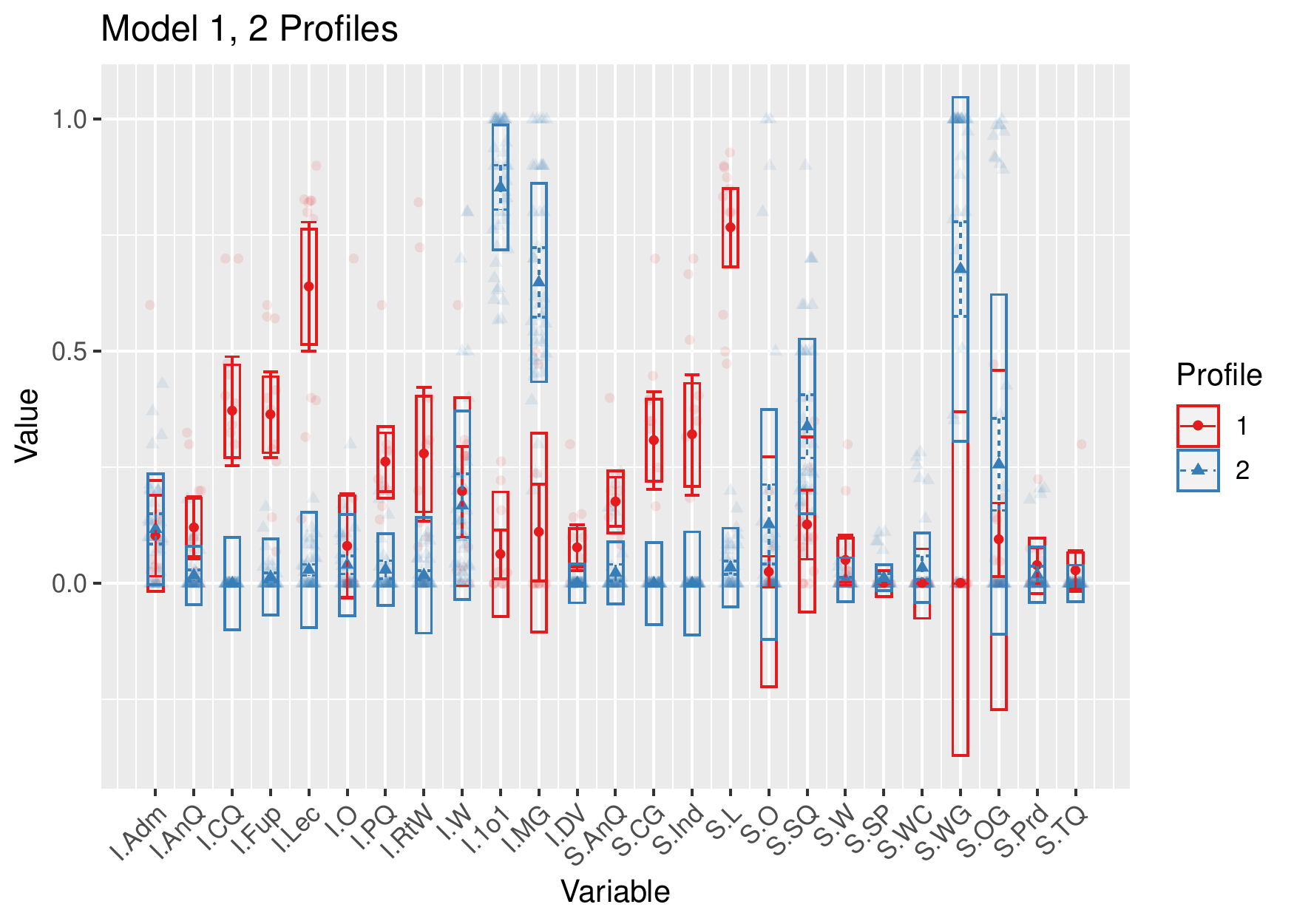}
\bigskip
    \includegraphics[height=0.4\textheight]{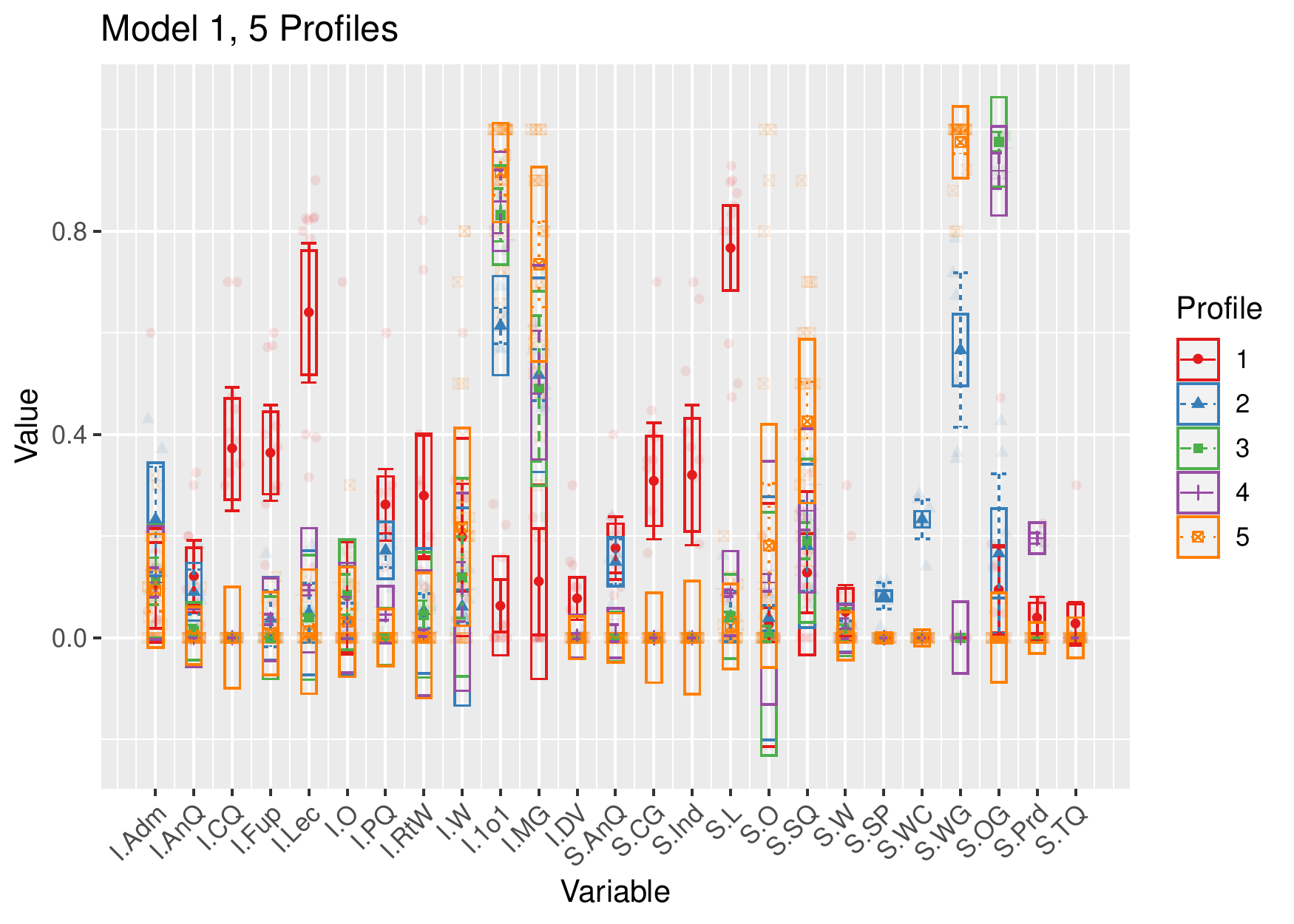}
    \caption{Model 1 shown for 2 profiles (top panel) and 5 profiles (bottom panel). The x-axis shows the full set of COPUS codes included in the analysis, while the y-axis shows the fraction of class time dedicated to that code. The light shaded dots show the actual COPUS profiles, colored by their assigned profile. The boxes show the standard deviation in each assigned profile, and the bars show the confidence interval of the centroid of said profile assignment. Since we used the equal variance model, the size of the boxes is equal between classes in each COPUS code.}
    \label{fig:FullList-Model1}
\end{figure*}

We ran LPA on our data using model 1 (equal variance between profiles, zero covariance) for 2-8 profiles. This range of profiles was originally chosen because it was the same range that is used in the TidyLPA vignette~\cite{tidylpa}, but was ultimately kept after investigating the results. The sorting of observations into profiles becomes nonsensical after 8 profiles. 
We present the 2 profile and 5 profile solutions here, and include the remainder in supplemental material. The 2 profile solution is included due to the successful sorting of observations into `interactive lecture-like' and `other', as seen in Table \ref{tab:2profiles}.  The top panel of figure \ref{fig:FullList-Model1} shows the box plots for each COPUS code and their associated profile groupings. 

The five profile solution was included because it had the best BIC with the lowest number of profiles. This fit successfully sorted the COPUS observations into `interactive lecture-like', Modeling Instruction, ISLE labs, CRP labs, and `recitation-like' categories, as seen in Table \ref{tab:5profiles}. The lower panel of figure \ref{fig:FullList-Model1} shows the box plots for each COPUS code and their associated profile groupings.



\begin{table*}[htbp]
\caption{Model 1 profile assignment, 2 profiles. Profile 1 is red on the top panel of figure \ref{fig:FullList-Model1}, while profile 2 is blue. ISLE had two sets of observations, the whole-class implementation (WC) and the lab-only implementation (LO).  }
\begin{ruledtabular}
\begin{tabular}{c|cccc}
  \textbf{Profile 1} & \multicolumn{4}{c}{\textbf{Profile 2}} \\
 \hline
Peer Instruction & ISLE LO Lab 1 & Modeling 1.1 & Tutorials 1 & Tutorials 11 \\
SCALE-UP 1.1 & ISLE LO Lab 2 & Modeling 1.2 & Tutorials 2 & Tutorials 12 \\
SCALE-UP 1.2 & ISLE LO Lab 3 & Modeling 2.1 & Tutorials 3 & Tutorials 13 \\
SCALE-UP 1.3 & ISLE LO Lab 4 & Modeling 2.2 & Tutorials 4 & Tutorials 14 \\
ISLE WC Lecture 1.1 & ISLE LO Lab 5 & Modeling 3.1 & Tutorials 5 & Tutorials 15 \\
ISLE WC Lecture 1.2 & ISLE WC Lab 1 & Modeling 3.2 & Tutorials 6 & Tutorials 16 \\
CRP 1.1 Lecture & ISLE WC Recitation 1 & CRP Discussion 1 & Tutorials 7 & Tutorials 17  \\
CRP 1.2 Lecture & ISLE WC Recitation 2 & CRP Discussion 2 & Tutorials 8 & Tutorials 18 \\
CRP 2.1 Lecture & ISLE WC Recitation 3 & CRP Lab 1 & Tutorials 9 & Tutorials 19 \\
CRP 2.2 Lecture & ISLE WC Recitation 4 & CRP Lab 2 & Tutorials 10  \\
Tutorials Lecture & & CRP Lab 3 \\
& & CRP Lab 4 
\label{tab:2profiles}
\end{tabular}
\end{ruledtabular}
\end{table*}

\begin{table*}[htbp]
\caption{Model 1 profile assignment, 5 profile. Profile 1 is red in the lower panel in figure \ref{fig:FullList-Model1}, profile 2 is blue, profile 3 is green, profile 4 is purple, and profile 5 is orange. ISLE had two sets of observations, the whole-class implementation (WC) and the lab-only implementation (LO).}
\begin{ruledtabular}
\begin{tabular}{c|c|c|c|c}
  \textbf{Profile 1} & \textbf{Profile 2} & \textbf{Profile 3} & \textbf{Profile 4} & \textbf{Profile 5} \\
 \hline
 Peer Instruction & Modeling 1.1 & ISLE WC Lab 1 & CRP Lab 1 & ISLE WC Recitation 1 \\
 
 SCALE-UP 1.1 & Modeling 1.2 & ISLE LO Lab 1 & CRP Lab 2    & ISLE WC Recitation 2 \\
 
 SCALE-UP 1.2 & Modeling 2.1 & ISLE LO Lab 2 & CRP Lab 3 & ISLE WC Recitation 3 \\
 
 SCALE-UP 1.3 & Modeling 2.2 & ISLE LO Lab 3 & CRP Lab 4 & ISLE WC Recitation 4 \\
 
 ISLE WC Lecture 1.1 & Modeling 3.1 & ISLE LO Lab 4 & & CRP Discussion 1 \\
 
 ISLE WC Lecture 1.2 & Modeling 3.2 & ISLE LO Lab 5 & & CRP Discussion 2 \\
 
 CRP 1.1 Lecture & & & & Tutorials 1 \\
 CRP 1.2 Lecture & & & & Tutorials 2  \\
 CRP 2.1 Lecture & & & & Tutorials 3 \\
 CRP 2.2 Lecture & & & & Tutorials 4 \\
 Tutorials Lecture & & & & Tutorials 5  \\
  & && & Tutorials 6  \\
  & && & Tutorials 7  \\    
  & && & Tutorials 8  \\
  & && & Tutorials 9  \\    
  & && & Tutorials 10 \\
  & && & Tutorials 11 \\    
  & && & Tutorials 12 \\
  & && & Tutorials 13 \\    
  & && & Tutorials 14 \\
  & && & Tutorials 15 \\    
  & && & Tutorials 16 \\
  & && & Tutorials 17 \\    
  & && & Tutorials 18 \\
  & && & Tutorials 19 
\label{tab:5profiles}
\end{tabular}
\end{ruledtabular}
\end{table*}

\begin{figure}[h]
    \includegraphics[width=\linewidth]{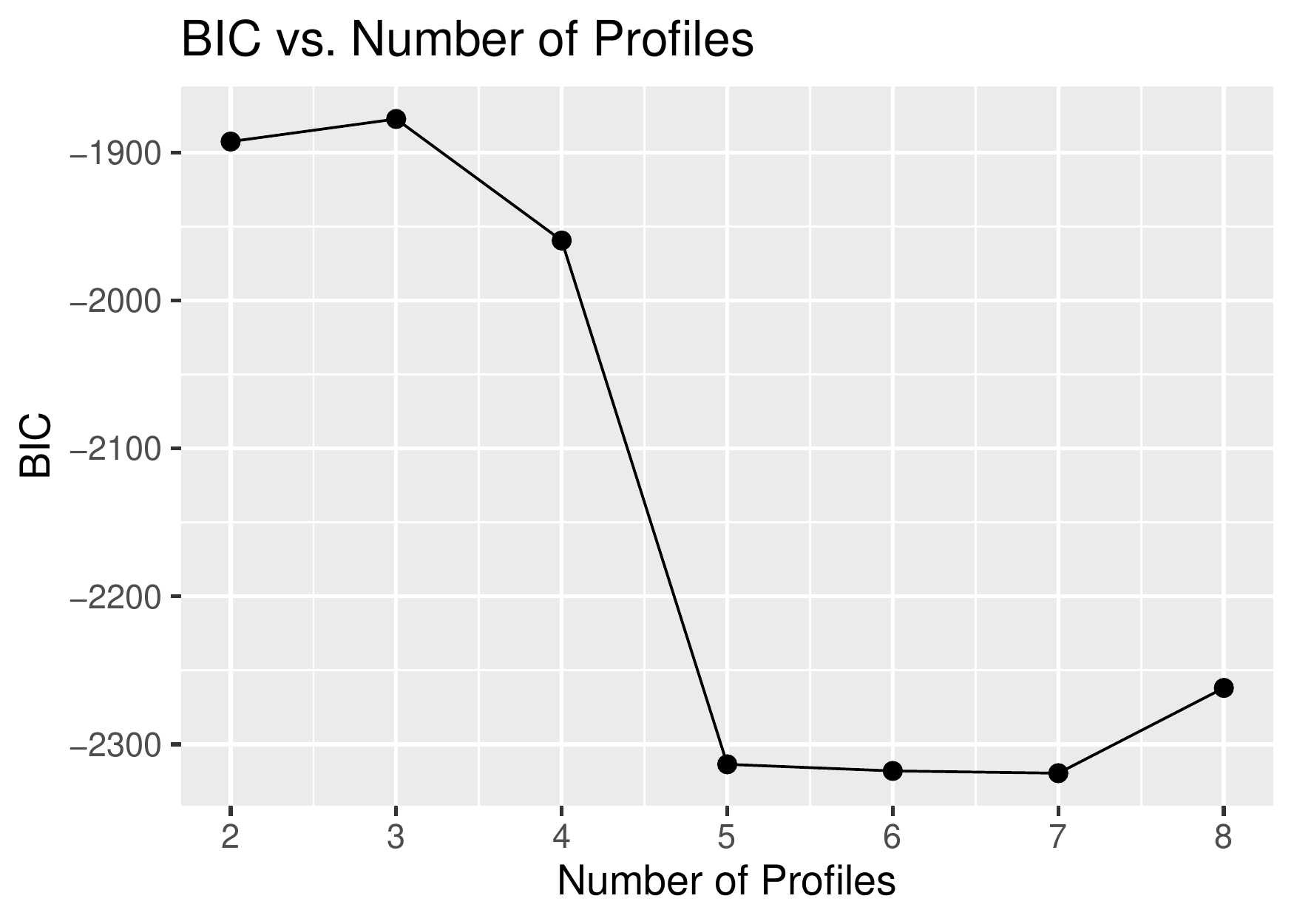}
    \caption{BIC vs number of profiles. We see a slight increase in BIC from two to three profiles, and then a sharp drop until five profiles. Five, six, and seven profiles have nearly equal BIC values, before increasing again with eight profiles. As such, we focus on two profiles and five profiles in further discussion.}
    \label{fig:BICModel1}
\end{figure}

The LPA algorithm uses hierarchical clustering to determine initial parameters, meaning the order of the data can have an effect on the resulting models.  Stains et. al. combated this by shuffling their data and running the analysis 10,000 times. As this paper is proof of concept with a limited data set, we randomized the data five times to determine if further randomization was needed. After 5 additional analyses, there were zero discrepancies between profile assignments. As such, we did not run further randomization trials.  
The Bayesian information criterion (BIC) was used as the primary indicator of model acceptability. In general, the further left on the number line, the better the fit of the model. The BIC is plotted for model 1 against the number of profiles that the COPUS data was sorted into, which can be seen in figure \ref{fig:BICModel1}. We see a slight increase in BIC from two to three profiles, and then a sharp drop until five profiles. Five, six, and seven profiles have nearly equal BIC values, before increasing again with eight profiles.

\section{Discussion}


\subsection{Two Profiles}
If we look more closely at the top panel of figure \ref{fig:FullList-Model1}, we can start to pick apart the differences in the assigned profiles. The two profiles that emerged effectively sorted the observations into `large lecture-like' and `other'. This becomes more apparent as we zoom into the individual codes. The largest separation between the two profiles occurs with the instructor lecturing (I.Lec), students listening (S.L), and the instructor having one on one conversations with students or small groups (I.1o1). If we think about how the observations were sorted (see table \ref{tab:2profiles} for the full list), this makes sense. While Peer instruction and SCALE-UP are pedagogies that incorporate active learning techniques, they are still typically large classes where the bulk of information transfer occurs. Pair those with the lecture components of the other pedagogies, and we can see how I.Lec and S.L were the starkest differences from the profile that encompasses lab and recitation components. The lecture-like profile that we see in our data more closely aligns with the hybrid category (``active lecture") that  ~\citet{Stains2018AnatomyUniversities} reports.

The code where profile 2 takes the lead, I.1o1, indicates that instructors spend more time engrossed in individual discussions with students outside of large lecture-like environments. This is to be expected when comparing small, typically small-group activity-centered environments to a large information transfer environment. The outlier to this generalization is Modeling Instruction; a large enrollment course with very little information transfer from the instructor via lecture, and large amounts of small group learning activities (which is likely why it was grouped with recitation and lab components).

On the other hand, there were a few codes that were nearly indistinguishable between the two profiles, including administrative tasks (I.Adm), students and instructor waiting (S.W and I.W), test/quiz (S.TQ), students making a prediction (S.Prd), and student presentations (S.SP). The only pedagogy that had student presentations was Modeling Instruction, when the students gathered for their `white board meetings'. Only one observation had a test or quiz, given in the lecture component of Tutorials. Student prediction is for when students are explicitly asked to make a prediction about the outcome of a demo or experiment, which was also a very rare occurrence in our set of observations. Meanwhile, administrative tasks, waiting around for students to finish an activity or the instructor to begin the next part of the lesson, and other unclassifiable activities (like eating a snack or taking a restroom break), seem to be universal to all pedagogies. 

Codes related to clicker questions or whole-class questioning were more common in the first profile, which is to be expected, as large enrollment courses are typically the environment for this instructional tool. These codes included I.CQ (instructor administers clicker question), I.Fup (instructor follows up a question with a longer discussion), I.PQ (instructor poses a non-rhetorical question to the whole class), S.AnQ (student answers a question with the whole class listening), S.CG (students work as a group to answer a clicker question), and S.Ind (students think independently to answer a question posed to the whole class).

While the first profile had very little group worksheet activity (S.WG) outside of SCALE-UP, the second profile had a large variance due to the wide array of group activities that are encompassed in these observations. Lab sections had mostly S.OG (other group activity), Tutorials had mostly S.WG, and Modeling Instruction had a mixture of both. As such, this code forced a large variance in the first profile as well, since we used model 1 during our latent profile analysis.


\subsection{Five Profiles}

Now let's look more closely at the bottom panel of figure \ref{fig:FullList-Model1}. Profile 1 (red boxes), which included lecture components and SCALE-UP observations, had a higher usage of clicker questions (I.CQ, S.CG, and S.Ind) and subsequent follow up (I.Fup). Profile 1 also had significantly higher usage of lecture (I.Lec) and thus students listening (S.L). Also noteworthy of profile 1 is a distinct lack of one-on-one discussions (I.1o1) and moving about the room to guide discussion (I.MG). 

Profile 2 (blue boxes), which holds all of the Modeling Instruction observations, is the only pedagogy that used student presentations (S.SP) and whole class discussions (S.WC). Profile 2 also shows slightly less time engaged in one-on-one discussions than the other active profiles. Less time was also spent engaged in group worksheets (S.WG) than the other active profiles.


Profiles 3 and 4, which hold the ISLE labs and the Context Rich Problem labs, respectively, are a little more difficult to distinguish. The most notable difference between these two lab sortings is the use of student predictions (S.Prd) in the CRP labs. In these observations, students had to predict the outcome of their experiments and discuss as a group before jumping into the activity, whereas the ISLE labs were less explicitly structured in this regard. The CRP labs were actually the first group to separate from the two-profile solution due to this code (see 3 profiles in the supplementary material), followed by the ISLE labs in the 4 profile solution. It is also worth mentioning that the ISLE labs all go together, even though most are the stand-alone implementation and one is part of the full-class implementation. That suggests that the lab-only implementation had high fidelity to the goals of the whole-class ISLE implementation.

Finally, the recitation-like observations in profile 5 stand out with their abundance of group worksheet activities (S.WG). This profile also has elevated levels of moving around the room to guide discussion (I.MG), as well as students raising their hands to ask questions (S.SQ).

\subsection{Further Discussion}
Modeling Instruction and SCALE-UP are both studio format with a mingled lecture-lab-recitation, but were not binned in the same profiles for any number of profile fits. Unlike Modeling Instruction, the SCALE-UP class we observed still used lecture as the primary means of information transfer, whereas the bulk of information transfer in a Modeling environment is through exploration by the students and the subsequent whole class discussions. This, plus the usage of clicker questions, was the biggest difference between the seemingly similar studio-based pedagogies. As mentioned in the introduction of the pedagogies, SCALE-UP is more-so descriptive of the physical environment rather than specific activities, and lends itself well to a combination of many styles of active learning. It is therefore possible that the COPUS profile of a SCALE-UP course could vary wildly between implementations, but it is also possible that a unique SCALE-UP signature could emerge that encompasses all codes due to the physical capability to cover all codes. More observations from several institutions will be necessary to investigate further.

It is also interesting to see how all of the recitation/discussion observations were grouped in both 2 and 5 profile cases. The largest distinction between these and the lab sections was the usage of S.WG and S.OG. The bulk of recitation/discussion section activities was with small group worksheets, hence the WG `working in groups on a worksheet' designation. Meanwhile, the labs were coded as `other group activity', as COPUS does not include a code for this purpose. Without this distinction, there would likely be a lot of overlap between lab and recitation profile assignment, as the base behaviors of working in small groups while the instructor mills about the room were largely similar. For future work, we recommend exploring different observation protocols that can better capture the difference between group activities, or using both COPUS and LOPUS (Lab Observation Protocol for Undergraduate STEM, which is the lab variant of COPUS)~\cite{Velasco2016CharacterizingSTEM}. 



\section{Conclusion}

We collected COPUS observations from six high-fidelity active learning pedagogies in physics. Using these observations, we performed latent profile analysis for 2-8 latent profile groupings. Two latent profiles successfully categorized the observations into `lecture-like' and `not lecture-like'. Five latent profiles sorted the observations into `lecture-like', Modeling Instruction, ISLE labs, Context-Rich labs, and recitation-like. With our preliminary data set, we made a first draft classification scheme using COPUS and LPA. 

This was a proof of concept study, which showed great promise for using LPA to study how COPUS profiles can be used to classify physics active learning pedagogies. Stains et. al. suggest that at least four COPUS observations are required to get a good snapshot of the class, because day to day fluctuations had some instructors categorized into more than one cluster in their study. While we were unable to achieve this with our current iteration of data collection, we suggest future studies take this suggestion into account, and make a `combined COPUS profile', consisting of several days worth of observations that include all components of a course (like lecure, lab, and recitation combined into one observation, as if a student were encountering the entire class). This method of data collection would allow us to more directly explore the difference between pedagogies, such as how a Peer Instruction experience with separate lab and recitation differs from a SCALE-UP integrated lab/lecture/recitation experience.

\bibliography{main} 

\begin{thebibliography}{26}%
\makeatletter
\providecommand \@ifxundefined [1]{%
 \@ifx{#1\undefined}
}%
\providecommand \@ifnum [1]{%
 \ifnum #1\expandafter \@firstoftwo
 \else \expandafter \@secondoftwo
 \fi
}%
\providecommand \@ifx [1]{%
 \ifx #1\expandafter \@firstoftwo
 \else \expandafter \@secondoftwo
 \fi
}%
\providecommand \natexlab [1]{#1}%
\providecommand \enquote  [1]{``#1''}%
\providecommand \bibnamefont  [1]{#1}%
\providecommand \bibfnamefont [1]{#1}%
\providecommand \citenamefont [1]{#1}%
\providecommand \href@noop [0]{\@secondoftwo}%
\providecommand \href [0]{\begingroup \@sanitize@url \@href}%
\providecommand \@href[1]{\@@startlink{#1}\@@href}%
\providecommand \@@href[1]{\endgroup#1\@@endlink}%
\providecommand \@sanitize@url [0]{\catcode `\\12\catcode `\$12\catcode
  `\&12\catcode `\#12\catcode `\^12\catcode `\_12\catcode `\%12\relax}%
\providecommand \@@startlink[1]{}%
\providecommand \@@endlink[0]{}%
\providecommand \url  [0]{\begingroup\@sanitize@url \@url }%
\providecommand \@url [1]{\endgroup\@href {#1}{\urlprefix }}%
\providecommand \urlprefix  [0]{URL }%
\providecommand \Eprint [0]{\href }%
\providecommand \doibase [0]{http://dx.doi.org/}%
\providecommand \selectlanguage [0]{\@gobble}%
\providecommand \bibinfo  [0]{\@secondoftwo}%
\providecommand \bibfield  [0]{\@secondoftwo}%
\providecommand \translation [1]{[#1]}%
\providecommand \BibitemOpen [0]{}%
\providecommand \bibitemStop [0]{}%
\providecommand \bibitemNoStop [0]{.\EOS\space}%
\providecommand \EOS [0]{\spacefactor3000\relax}%
\providecommand \BibitemShut  [1]{\csname bibitem#1\endcsname}%
\let\auto@bib@innerbib\@empty
\bibitem [{\citenamefont {Meltzer}\ and\ \citenamefont
  {Thornton}(2012)}]{Meltzer2012ResourcePhysics}%
  \BibitemOpen
  \bibfield  {author} {\bibinfo {author} {\bibfnamefont {D.~E.}\ \bibnamefont
  {Meltzer}}\ and\ \bibinfo {author} {\bibfnamefont {R.~K.}\ \bibnamefont
  {Thornton}},\ }\href {\doibase 10.1119/1.3678299} {\bibfield  {journal}
  {\bibinfo  {journal} {Am. J. Phys.}\ }\textbf {\bibinfo {volume} {80}},\
  \bibinfo {pages} {478} (\bibinfo {year} {2012})}\BibitemShut {NoStop}%
\bibitem [{\citenamefont {Redish}\ and\ \citenamefont
  {Cooney}(2007)}]{redish_research-based_2007}%
  \BibitemOpen
  \bibinfo {editor} {\bibfnamefont {E.~F.}\ \bibnamefont {Redish}}\ and\
  \bibinfo {editor} {\bibfnamefont {P.~J.}\ \bibnamefont {Cooney}},\ eds.,\
  \href {https://www.compadre.org/per/per_reviews/volume1.cfm} {\emph {\bibinfo
  {title} {Research-{Based} {Reform} of {University} {Physics}}}},\ \bibinfo
  {series} {Reviews in {PER}}, Vol.~\bibinfo {volume} {1}\ (\bibinfo
  {publisher} {American Association of Physics Teachers},\ \bibinfo {year}
  {2007})\BibitemShut {NoStop}%
\bibitem [{\citenamefont {Thacker}(2003)}]{thacker_recent_2003}%
  \BibitemOpen
  \bibfield  {author} {\bibinfo {author} {\bibfnamefont {B.~A.}\ \bibnamefont
  {Thacker}},\ }\href {\doibase 10.1088/0034-4885/66/10/R07} {\bibfield
  {journal} {\bibinfo  {journal} {Reports on Progress in Physics}\ }\textbf
  {\bibinfo {volume} {66}},\ \bibinfo {pages} {1833} (\bibinfo {year}
  {2003})}\BibitemShut {NoStop}%
\bibitem [{\citenamefont {Commeford}\ \emph {et~al.}(2020)\citenamefont
  {Commeford}, \citenamefont {Brewe},\ and\ \citenamefont
  {Traxler}}]{Commeford2020CharacterizingObservations}%
  \BibitemOpen
  \bibfield  {author} {\bibinfo {author} {\bibfnamefont {K.}~\bibnamefont
  {Commeford}}, \bibinfo {author} {\bibfnamefont {E.}~\bibnamefont {Brewe}}, \
  and\ \bibinfo {author} {\bibfnamefont {A.}~\bibnamefont {Traxler}},\
  }\href@noop {} {\emph {\bibinfo {title} {{Characterizing active learning
  environments in physics using network analysis and COPUS observations}}}},\
  \bibinfo {type} {Tech. Rep.}\ (\bibinfo {year} {2020})\BibitemShut {NoStop}%
\bibitem [{\citenamefont {Smith}\ \emph {et~al.}(2013)\citenamefont {Smith},
  \citenamefont {Jones}, \citenamefont {Gilbert},\ and\ \citenamefont
  {Wieman}}]{Smith2013}%
  \BibitemOpen
  \bibfield  {author} {\bibinfo {author} {\bibfnamefont {M.~K.}\ \bibnamefont
  {Smith}}, \bibinfo {author} {\bibfnamefont {F.~H.~M.}\ \bibnamefont {Jones}},
  \bibinfo {author} {\bibfnamefont {S.~L.}\ \bibnamefont {Gilbert}}, \ and\
  \bibinfo {author} {\bibfnamefont {C.~E.}\ \bibnamefont {Wieman}},\ }\href
  {\doibase 10.1187/cbe.13-08-0154} {\bibfield  {journal} {\bibinfo  {journal}
  {CBE Life Sci. Educ.}\ }\textbf {\bibinfo {volume} {12}},\ \bibinfo {pages}
  {618} (\bibinfo {year} {2013})}\BibitemShut {NoStop}%
\bibitem [{\citenamefont {Traxler}\ \emph {et~al.}(2020)\citenamefont
  {Traxler}, \citenamefont {Suda}, \citenamefont {Brewe},\ and\ \citenamefont
  {Commeford}}]{Traxler2020NetworkPhysics}%
  \BibitemOpen
  \bibfield  {author} {\bibinfo {author} {\bibfnamefont {A.~L.}\ \bibnamefont
  {Traxler}}, \bibinfo {author} {\bibfnamefont {T.}~\bibnamefont {Suda}},
  \bibinfo {author} {\bibfnamefont {E.}~\bibnamefont {Brewe}}, \ and\ \bibinfo
  {author} {\bibfnamefont {K.}~\bibnamefont {Commeford}},\ }\href {\doibase
  10.1103/PhysRevPhysEducRes.16.020129} {\bibfield  {journal} {\bibinfo
  {journal} {Physical Review Physics Education Research}\ }\textbf {\bibinfo
  {volume} {16}},\ \bibinfo {pages} {020129} (\bibinfo {year}
  {2020})}\BibitemShut {NoStop}%
\bibitem [{\citenamefont {Stains}\ \emph {et~al.}(2018)\citenamefont {Stains},
  \citenamefont {Harshman}, \citenamefont {Barker}, \citenamefont {Chasteen},
  \citenamefont {Cole}, \citenamefont {DeChenne-Peters}, \citenamefont {Eagan},
  \citenamefont {Esson}, \citenamefont {Knight}, \citenamefont {Laski},
  \citenamefont {Levis-Fitzgerald}, \citenamefont {Lee}, \citenamefont {Lo},
  \citenamefont {McDonnell}, \citenamefont {McKay}, \citenamefont {Michelotti},
  \citenamefont {Musgrove}, \citenamefont {Palmer}, \citenamefont {Plank},
  \citenamefont {Rodela}, \citenamefont {Sanders}, \citenamefont {Schimpf},
  \citenamefont {Schulte}, \citenamefont {Smith}, \citenamefont {Stetzer},
  \citenamefont {Van~Valkenburgh}, \citenamefont {Vinson}, \citenamefont
  {Weir}, \citenamefont {Wendel}, \citenamefont {Wheeler},\ and\ \citenamefont
  {Young}}]{Stains2018AnatomyUniversities}%
  \BibitemOpen
  \bibfield  {author} {\bibinfo {author} {\bibfnamefont {M.}~\bibnamefont
  {Stains}}, \bibinfo {author} {\bibfnamefont {J.}~\bibnamefont {Harshman}},
  \bibinfo {author} {\bibfnamefont {M.~K.}\ \bibnamefont {Barker}}, \bibinfo
  {author} {\bibfnamefont {S.~V.}\ \bibnamefont {Chasteen}}, \bibinfo {author}
  {\bibfnamefont {R.}~\bibnamefont {Cole}}, \bibinfo {author} {\bibfnamefont
  {S.~E.}\ \bibnamefont {DeChenne-Peters}}, \bibinfo {author} {\bibfnamefont
  {M.~K.}\ \bibnamefont {Eagan}}, \bibinfo {author} {\bibfnamefont {J.~M.}\
  \bibnamefont {Esson}}, \bibinfo {author} {\bibfnamefont {J.~K.}\ \bibnamefont
  {Knight}}, \bibinfo {author} {\bibfnamefont {F.~A.}\ \bibnamefont {Laski}},
  \bibinfo {author} {\bibfnamefont {M.}~\bibnamefont {Levis-Fitzgerald}},
  \bibinfo {author} {\bibfnamefont {C.~J.}\ \bibnamefont {Lee}}, \bibinfo
  {author} {\bibfnamefont {S.~M.}\ \bibnamefont {Lo}}, \bibinfo {author}
  {\bibfnamefont {L.~M.}\ \bibnamefont {McDonnell}}, \bibinfo {author}
  {\bibfnamefont {T.~A.}\ \bibnamefont {McKay}}, \bibinfo {author}
  {\bibfnamefont {N.}~\bibnamefont {Michelotti}}, \bibinfo {author}
  {\bibfnamefont {A.}~\bibnamefont {Musgrove}}, \bibinfo {author}
  {\bibfnamefont {M.~S.}\ \bibnamefont {Palmer}}, \bibinfo {author}
  {\bibfnamefont {K.~M.}\ \bibnamefont {Plank}}, \bibinfo {author}
  {\bibfnamefont {T.~M.}\ \bibnamefont {Rodela}}, \bibinfo {author}
  {\bibfnamefont {E.~R.}\ \bibnamefont {Sanders}}, \bibinfo {author}
  {\bibfnamefont {N.~G.}\ \bibnamefont {Schimpf}}, \bibinfo {author}
  {\bibfnamefont {P.~M.}\ \bibnamefont {Schulte}}, \bibinfo {author}
  {\bibfnamefont {M.~K.}\ \bibnamefont {Smith}}, \bibinfo {author}
  {\bibfnamefont {M.}~\bibnamefont {Stetzer}}, \bibinfo {author} {\bibfnamefont
  {B.}~\bibnamefont {Van~Valkenburgh}}, \bibinfo {author} {\bibfnamefont
  {E.}~\bibnamefont {Vinson}}, \bibinfo {author} {\bibfnamefont {L.~K.}\
  \bibnamefont {Weir}}, \bibinfo {author} {\bibfnamefont {P.~J.}\ \bibnamefont
  {Wendel}}, \bibinfo {author} {\bibfnamefont {L.~B.}\ \bibnamefont {Wheeler}},
  \ and\ \bibinfo {author} {\bibfnamefont {A.~M.}\ \bibnamefont {Young}},\
  }\href {\doibase 10.1126/science.aap8892} {\bibfield  {journal} {\bibinfo
  {journal} {Science}\ }\textbf {\bibinfo {volume} {359}},\ \bibinfo {pages}
  {1468} (\bibinfo {year} {2018})}\BibitemShut {NoStop}%
\bibitem [{\citenamefont {McDermott}\ and\ \citenamefont
  {Shaffer}(2002)}]{McDermott2002}%
  \BibitemOpen
  \bibfield  {author} {\bibinfo {author} {\bibfnamefont {L.~C.}\ \bibnamefont
  {McDermott}}\ and\ \bibinfo {author} {\bibfnamefont {P.~S.}\ \bibnamefont
  {Shaffer}},\ }\href@noop {} {\emph {\bibinfo {title} {{Tutorials in
  Introductory Physics}}}},\ \bibinfo {edition} {1st}\ ed.\ (\bibinfo
  {publisher} {Prentice Hall},\ \bibinfo {address} {Upper Saddle River, NJ},\
  \bibinfo {year} {2002})\BibitemShut {NoStop}%
\bibitem [{\citenamefont {Scherr}\ and\ \citenamefont
  {Elby}(2007)}]{Scherr2007EnablingMaterials}%
  \BibitemOpen
  \bibfield  {author} {\bibinfo {author} {\bibfnamefont {R.~E.}\ \bibnamefont
  {Scherr}}\ and\ \bibinfo {author} {\bibfnamefont {A.}~\bibnamefont {Elby}},\
  }\href {\doibase 10.1063/1.2508688} {\bibfield  {journal} {\bibinfo
  {journal} {AIP Conf. Proc.}\ }\textbf {\bibinfo {volume} {883}},\ \bibinfo
  {pages} {46} (\bibinfo {year} {2007})}\BibitemShut {NoStop}%
\bibitem [{\citenamefont {Etkina}\ and\ \citenamefont
  {Heuvelen}(2007)}]{Etkina2007InvestigativePhysics}%
  \BibitemOpen
  \bibfield  {author} {\bibinfo {author} {\bibfnamefont {E.}~\bibnamefont
  {Etkina}}\ and\ \bibinfo {author} {\bibfnamefont {A.~V.}\ \bibnamefont
  {Heuvelen}},\ }in\ \href@noop {} {\emph {\bibinfo {booktitle} {Reviews in
  PER: Research-Based Reform of University Physics}}},\ Vol.~\bibinfo {volume}
  {1},\ \bibinfo {editor} {edited by\ \bibinfo {editor} {\bibfnamefont {E.~F.}\
  \bibnamefont {Redish}}\ and\ \bibinfo {editor} {\bibfnamefont {P.~J.}\
  \bibnamefont {Cooney}}}\ (\bibinfo  {publisher} {American Association of
  Physics Teachers},\ \bibinfo {address} {College Park, MD},\ \bibinfo {year}
  {2007})\BibitemShut {NoStop}%
\bibitem [{\citenamefont {Etkina}(2015)}]{Etkina2015MillikanPractices}%
  \BibitemOpen
  \bibfield  {author} {\bibinfo {author} {\bibfnamefont {E.}~\bibnamefont
  {Etkina}},\ }\href {\doibase 10.1119/1.4923432} {\bibfield  {journal}
  {\bibinfo  {journal} {Am. J. Phys.}\ }\textbf {\bibinfo {volume} {83}},\
  \bibinfo {pages} {669} (\bibinfo {year} {2015})}\BibitemShut {NoStop}%
\bibitem [{Phy()}]{PhysPortInstruction}%
  \BibitemOpen
  \href
  {https://www.physport.org/methods/method.cfm?G=University_Modeling_Instruction}
  {\enquote {\bibinfo {title} {{PhysPort Methods and Materials: University
  Modeling Instruction}},}\ }\BibitemShut {NoStop}%
\bibitem [{\citenamefont {Wells}\ \emph {et~al.}(1995)\citenamefont {Wells},
  \citenamefont {Hestenes},\ and\ \citenamefont
  {Swackhamer}}]{Wells1995AInstruction}%
  \BibitemOpen
  \bibfield  {author} {\bibinfo {author} {\bibfnamefont {M.}~\bibnamefont
  {Wells}}, \bibinfo {author} {\bibfnamefont {D.}~\bibnamefont {Hestenes}}, \
  and\ \bibinfo {author} {\bibfnamefont {G.}~\bibnamefont {Swackhamer}},\
  }\href {\doibase 10.1119/1.17849} {\bibfield  {journal} {\bibinfo  {journal}
  {Am. J. Phys.}\ }\textbf {\bibinfo {volume} {63}},\ \bibinfo {pages} {606}
  (\bibinfo {year} {1995})}\BibitemShut {NoStop}%
\bibitem [{\citenamefont {Brewe}(2008)}]{Brewe2008}%
  \BibitemOpen
  \bibfield  {author} {\bibinfo {author} {\bibfnamefont {E.}~\bibnamefont
  {Brewe}},\ }\href {\doibase 10.1119/1.2983148} {\bibfield  {journal}
  {\bibinfo  {journal} {Am. J. Phys.}\ }\textbf {\bibinfo {volume} {76}},\
  \bibinfo {pages} {1155} (\bibinfo {year} {2008})}\BibitemShut {NoStop}%
\bibitem [{\citenamefont {Mazur}(1997)}]{Mazur1997}%
  \BibitemOpen
  \bibfield  {author} {\bibinfo {author} {\bibfnamefont {E.}~\bibnamefont
  {Mazur}},\ }\href@noop {} {\emph {\bibinfo {title} {{Peer Instruction: A
  User's Manual}}}},\ \bibinfo {edition} {1st}\ ed.\ (\bibinfo  {publisher}
  {Prentice Hall},\ \bibinfo {address} {Upper Saddle River, NJ},\ \bibinfo
  {year} {1997})\BibitemShut {NoStop}%
\bibitem [{\citenamefont {Crouch}\ \emph {et~al.}(2007)\citenamefont {Crouch},
  \citenamefont {Watkins}, \citenamefont {Fagen},\ and\ \citenamefont
  {Mazur}}]{Crouch2007PeerOnce}%
  \BibitemOpen
  \bibfield  {author} {\bibinfo {author} {\bibfnamefont {C.}~\bibnamefont
  {Crouch}}, \bibinfo {author} {\bibfnamefont {J.}~\bibnamefont {Watkins}},
  \bibinfo {author} {\bibfnamefont {A.}~\bibnamefont {Fagen}}, \ and\ \bibinfo
  {author} {\bibfnamefont {E.}~\bibnamefont {Mazur}},\ }in\ \href@noop {}
  {\emph {\bibinfo {booktitle} {Reviews in PER: Research-Based Reform of
  University Physics}}},\ Vol.~\bibinfo {volume} {1},\ \bibinfo {editor}
  {edited by\ \bibinfo {editor} {\bibfnamefont {E.~F.}\ \bibnamefont {Redish}}\
  and\ \bibinfo {editor} {\bibfnamefont {P.~J.}\ \bibnamefont {Cooney}}}\
  (\bibinfo  {publisher} {American Association of Physics Teachers},\ \bibinfo
  {address} {College Park, MD},\ \bibinfo {year} {2007})\BibitemShut {NoStop}%
\bibitem [{Uni()}]{UniversityDevelopment}%
  \BibitemOpen
  \href {http://groups.physics.umn.edu/physed/Research/MNModel/MMt.html}
  {\enquote {\bibinfo {title} {{University of Minnesota Physics Education
  Research and Development: Minnesota Model for Large Introductory Courses}},}\
  }\BibitemShut {NoStop}%
\bibitem [{\citenamefont {Heller}\ and\ \citenamefont
  {Hollabaugh}(1992)}]{Heller1992TeachingGroups}%
  \BibitemOpen
  \bibfield  {author} {\bibinfo {author} {\bibfnamefont {P.}~\bibnamefont
  {Heller}}\ and\ \bibinfo {author} {\bibfnamefont {M.}~\bibnamefont
  {Hollabaugh}},\ }\href {\doibase 10.1119/1.17118} {\bibfield  {journal}
  {\bibinfo  {journal} {Am. J. Phys.}\ }\textbf {\bibinfo {volume} {60}},\
  \bibinfo {pages} {637} (\bibinfo {year} {1992})}\BibitemShut {NoStop}%
\bibitem [{\citenamefont {Beichner}\ \emph {et~al.}(2007)\citenamefont
  {Beichner}, \citenamefont {Saul}, \citenamefont {Abbott}, \citenamefont
  {Morse}, \citenamefont {Deardorff}, \citenamefont {Allain}, \citenamefont
  {Bonham}, \citenamefont {Dancy},\ and\ \citenamefont
  {Risley}}]{Beichner2007TheProject}%
  \BibitemOpen
  \bibfield  {author} {\bibinfo {author} {\bibfnamefont {R.~J.}\ \bibnamefont
  {Beichner}}, \bibinfo {author} {\bibfnamefont {J.~M.}\ \bibnamefont {Saul}},
  \bibinfo {author} {\bibfnamefont {D.~S.}\ \bibnamefont {Abbott}}, \bibinfo
  {author} {\bibfnamefont {J.~J.}\ \bibnamefont {Morse}}, \bibinfo {author}
  {\bibfnamefont {D.}~\bibnamefont {Deardorff}}, \bibinfo {author}
  {\bibfnamefont {R.~J.}\ \bibnamefont {Allain}}, \bibinfo {author}
  {\bibfnamefont {S.~W.}\ \bibnamefont {Bonham}}, \bibinfo {author}
  {\bibfnamefont {M.~H.}\ \bibnamefont {Dancy}}, \ and\ \bibinfo {author}
  {\bibfnamefont {J.~S.}\ \bibnamefont {Risley}},\ }in\ \href
  {http://www.percentral.com/PER/per_reviews/media/volume1/SCALE-UP-2007.pdf}
  {\emph {\bibinfo {booktitle} {Reviews in PER: Research-Based Reform of
  University Physics}}},\ Vol.~\bibinfo {volume} {1},\ \bibinfo {editor}
  {edited by\ \bibinfo {editor} {\bibfnamefont {E.~F.}\ \bibnamefont {Redish}}\
  and\ \bibinfo {editor} {\bibfnamefont {P.~J.}\ \bibnamefont {Cooney}}}\
  (\bibinfo  {publisher} {American Association of Physics Teachers},\ \bibinfo
  {address} {College Park, MD},\ \bibinfo {year} {2007})\BibitemShut {NoStop}%
\bibitem [{\citenamefont {Oberski}(2016)}]{Oberski2016MixtureAnalysis}%
  \BibitemOpen
  \bibfield  {author} {\bibinfo {author} {\bibfnamefont {D.}~\bibnamefont
  {Oberski}},\ }in\ \href {\doibase 10.1007/978-3-319-26633-6{\_}12} {\emph
  {\bibinfo {booktitle} {Modern statistical methods for HCI}}},\ \bibinfo
  {editor} {edited by\ \bibinfo {editor} {\bibfnamefont {J.}~\bibnamefont
  {Robertson}}\ and\ \bibinfo {editor} {\bibfnamefont {M.}~\bibnamefont
  {Kaptein}}}\ (\bibinfo  {publisher} {Springer International Publishing},\
  \bibinfo {address} {Switzerland},\ \bibinfo {year} {2016})\ pp.\ \bibinfo
  {pages} {275--287}\BibitemShut {NoStop}%
\bibitem [{\citenamefont {Stains}\ and\ \citenamefont
  {Harshman}()}]{StainsCOPUSAnalyzer}%
  \BibitemOpen
  \bibfield  {author} {\bibinfo {author} {\bibfnamefont {M.}~\bibnamefont
  {Stains}}\ and\ \bibinfo {author} {\bibfnamefont {J.}~\bibnamefont
  {Harshman}},\ }\href@noop {} {\enquote {\bibinfo {title} {{COPUS
  Analyzer}},}\ }\BibitemShut {NoStop}%
\bibitem [{Note1()}]{Note1}%
  \BibitemOpen
  \bibinfo {note} {Victor Lavrenko does in incredible job illustrating how this
  process works on his YouTube channel http://bit.ly/EM-alg}\BibitemShut
  {NoStop}%
\bibitem [{\citenamefont {Fraley}\ and\ \citenamefont
  {Raftery}(2007)}]{Fraley2007BayesianClustering}%
  \BibitemOpen
  \bibfield  {author} {\bibinfo {author} {\bibfnamefont {C.}~\bibnamefont
  {Fraley}}\ and\ \bibinfo {author} {\bibfnamefont {A.~E.}\ \bibnamefont
  {Raftery}},\ }\href {\doibase 10.1007/s00357-007-0004-5} {\bibfield
  {journal} {\bibinfo  {journal} {Journal of Classification}\ }\textbf
  {\bibinfo {volume} {24}},\ \bibinfo {pages} {155} (\bibinfo {year}
  {2007})}\BibitemShut {NoStop}%
\bibitem [{\citenamefont {Rosenberg}\ \emph {et~al.}(2018)\citenamefont
  {Rosenberg}, \citenamefont {Beymer}, \citenamefont {Anderson}, \citenamefont
  {{Van Lissa}},\ and\ \citenamefont {Schmidt}}]{tidylpa}%
  \BibitemOpen
  \bibfield  {author} {\bibinfo {author} {\bibfnamefont {J.~M.}\ \bibnamefont
  {Rosenberg}}, \bibinfo {author} {\bibfnamefont {P.~N.}\ \bibnamefont
  {Beymer}}, \bibinfo {author} {\bibfnamefont {D.~J.}\ \bibnamefont
  {Anderson}}, \bibinfo {author} {\bibfnamefont {C.~J.}\ \bibnamefont {{Van
  Lissa}}}, \ and\ \bibinfo {author} {\bibfnamefont {J.~A.}\ \bibnamefont
  {Schmidt}},\ }\href {\doibase 10.21105/joss.00978} {\bibfield  {journal}
  {\bibinfo  {journal} {Journal of Open Source Software}\ }\textbf {\bibinfo
  {volume} {3}},\ \bibinfo {pages} {978} (\bibinfo {year} {2018})}\BibitemShut
  {NoStop}%
\bibitem [{\citenamefont {Muth{\'{e}}n}\ and\ \citenamefont
  {Muth{\'{e}}n}(1998)}]{Muthen1998MplusGuide}%
  \BibitemOpen
  \bibfield  {author} {\bibinfo {author} {\bibfnamefont {L.~K.}\ \bibnamefont
  {Muth{\'{e}}n}}\ and\ \bibinfo {author} {\bibfnamefont {B.~O.}\ \bibnamefont
  {Muth{\'{e}}n}},\ }\href {www.StatModel.com} {\enquote {\bibinfo {title}
  {{Mplus User's Guide}},}\ } (\bibinfo {year} {1998})\BibitemShut {NoStop}%
\bibitem [{\citenamefont {Velasco}\ \emph {et~al.}(2016)\citenamefont
  {Velasco}, \citenamefont {Knedeisen}, \citenamefont {Xue}, \citenamefont
  {Vickrey}, \citenamefont {Abebe},\ and\ \citenamefont
  {Stains}}]{Velasco2016CharacterizingSTEM}%
  \BibitemOpen
  \bibfield  {author} {\bibinfo {author} {\bibfnamefont {J.~B.}\ \bibnamefont
  {Velasco}}, \bibinfo {author} {\bibfnamefont {A.}~\bibnamefont {Knedeisen}},
  \bibinfo {author} {\bibfnamefont {D.}~\bibnamefont {Xue}}, \bibinfo {author}
  {\bibfnamefont {T.~L.}\ \bibnamefont {Vickrey}}, \bibinfo {author}
  {\bibfnamefont {M.}~\bibnamefont {Abebe}}, \ and\ \bibinfo {author}
  {\bibfnamefont {M.}~\bibnamefont {Stains}},\ }\href {\doibase
  10.1021/acs.jchemed.6b00062} {\bibfield  {journal} {\bibinfo  {journal}
  {Journal of Chemical Education}\ }\textbf {\bibinfo {volume} {93}},\ \bibinfo
  {pages} {1191} (\bibinfo {year} {2016})}\BibitemShut {NoStop}%
\end{thebibliography}%

\end{document}


\title{Supplemental Material}
\maketitle

In this supplementary material, we include the box plots and grouping outputs for all but the 2 and 5 profile solutions from latent profile analysis, which are discussed in the main paper. The majority of discussion will be contained in the captions, to ensure physical proximity to the figures in question, so the reader does not have to flip back and forth between the text and the figures/tables. We also include the table of COPUS codes.

\begin{table*}
\begin{ruledtabular}
\caption{Model 1, 3 profiles. This output kept the sorting from the 2 profile solution discussed in the main text, but pulled the Context-Rich problems labs into their own category.}
\begin{tabular}{ccc}
  Class 1    & Class 2  & Class 3      \\
 \hline
 Peer Instruction    &  Modeling 1.1  & CRP Lab 1    \\
 SCALE-UP 1.1        &  Modeling 1.2  & CRP Lab 2      \\
 SCALE-UP 1.2        &  Modeling 2.1  & CRP Lab 3     \\
 SCALE-UP 1.3        &  Modeling 2.2  & CRP Lab 4     \\
 ISLE WC Lecture 1.1 & Modeling 3.1   \\
 ISLE WC Lecture 1.2 & Modeling 3.2  \\
 CRP Lecture 1.1     & ISLE WC Lab 1  \\
 CRP Lecture 1.2     & ISLE WC Recitation 1 \\
 CRP Lecture 2.1     & ISLE WC Recitation 2\\
 CRP Lecture 2.2     & ISLE WC Recitation 3  \\
 Tutorials Lecture   & ISLE WC Recitation 4 \\
                     & ISLE LO Lab 1 \\
                     & ISLE LO Lab 2 \\   
                     & ISLE LO Lab 3 \\
                     & ISLE LO Lab 4 \\    
                     & ISLE LO Lab 5 \\
                     & CRP Discussion 1 \\
                     & CRP Discussion 2 \\
                     & Tutorials 1            \\    
                     & Tutorials 2               \\
                     & Tutorials 3           \\    
                     & Tutorials 4              \\
                     & Tutorials 5           \\    
                     & Tutorials 6              \\
                     & Tutorials 7           \\    
                     & Tutorials 8              \\
                     & Tutorials 9           \\    
                     & Tutorials 10              \\
                     & Tutorials 11           \\    
                     & Tutorials 12              \\
                     & Tutorials 13           \\    
                     & Tutorials 14              \\
                     & Tutorials 15           \\    
                     & Tutorials 16              \\
                     & Tutorials 17           \\    
                     & Tutorials 18              \\
                     & Tutorials 19  \\
\end{tabular}
\end{ruledtabular}
\end{table*}

\begin{figure*}
    \includegraphics[width=\linewidth]{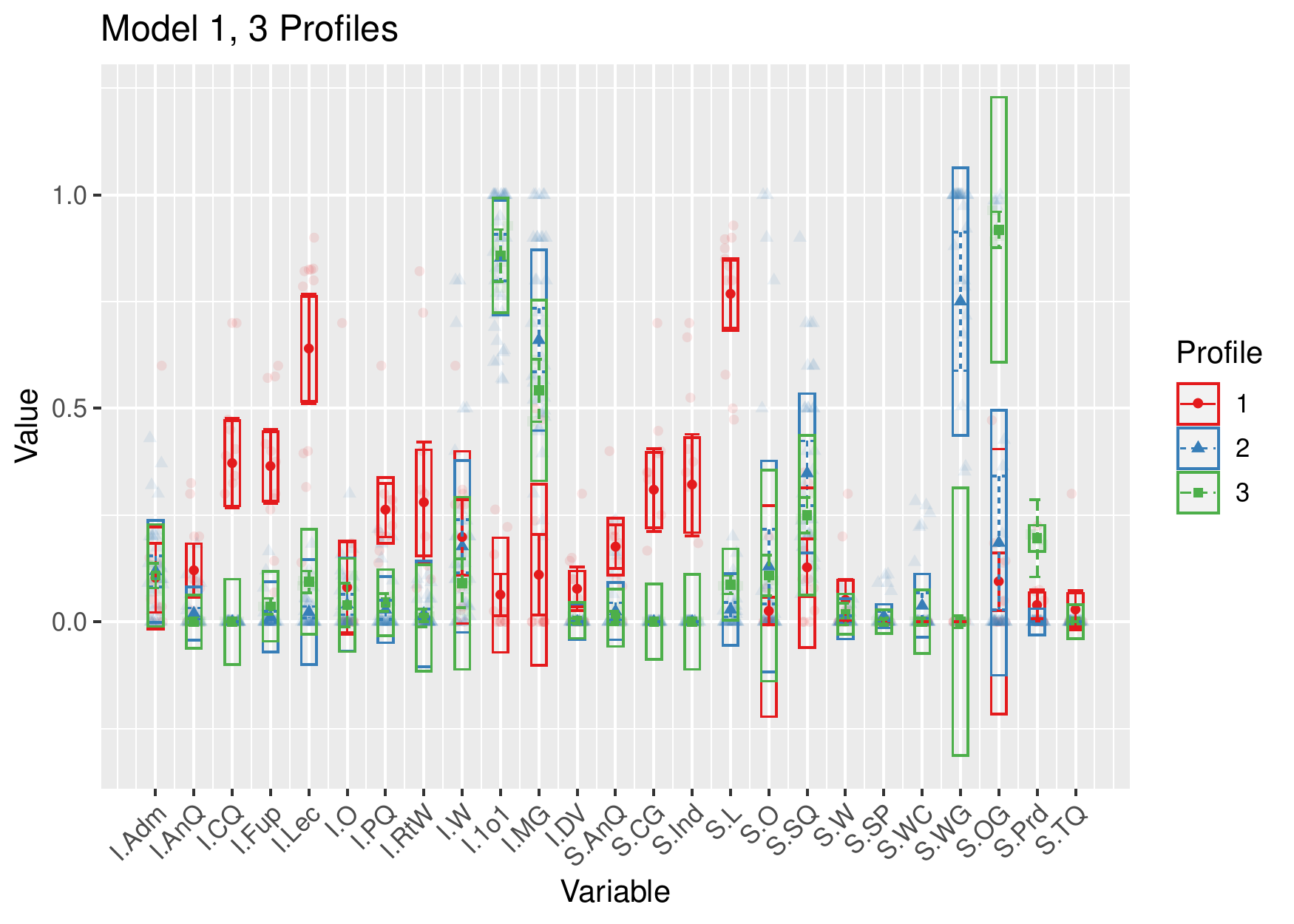}
    \caption{Model 1, 3 profiles. This output kept the sorting from the 2 profile solution presented in the main text, but pulled the Context-Rich problems labs into their own category (profile 3, in green). The x-axis shows the full set of COPUS codes included in the analysis, while the y-axis shows the fraction of class time dedicated to that code. The light shaded dots show the actual COPUS profiles, colored by their assigned profile. The boxes show the standard deviation in each assigned profile, and the bars show the confidence interval of the centroid. Since we used the equal variance model, the size of the boxes is equal between classes in each COPUS code. }
    \label{fig:FullList-Model1-3profiles}
\end{figure*}

\begin{table*}[p!]
\begin{ruledtabular}
\caption{Model 1, 4 profiles. This output kept the sorting from the 3 profile solution, but pulled the ISLE labs into their own category. }
\begin{tabular}{cccc}
  Class 1    & Class 2  & Class 3 & Class 4     \\
 \hline
 Peer Instruction    & Modeling 1.1         & ISLE WC Lab 1 & CRP Lab 1      \\
 SCALE-UP 1.1        & Modeling 1.2         & ISLE LO Lab 1 & CRP Lab 2       \\
 SCALE-UP 1.2        & Modeling 2.1         & ISLE LO Lab 2 & CRP Lab 3      \\
 SCALE-UP 1.3        & Modeling 2.2         & ISLE LO Lab 3 & CRP Lab 4      \\
 ISLE WC Lecture 1.1 & Modeling 3.1         & ISLE LO Lab 4 \\
 ISLE WC Lecture 1.2 & Modeling 3.2         & ISLE LO Lab 5\\
 CRP Lecture 1.1     & ISLE WC Recitation 1 \\
 CRP Lecture 1.2     & ISLE WC Recitation 2 \\
 CRP Lecture 2.1     & ISLE WC Recitation 3 \\
 CRP Lecture 2.2     & ISLE WC Recitation 4 \\
 Tutorials Lecture   & CRP Discussion 1 \\
                     & CRP Discussion 2 \\
                     & Tutorials 1     \\    
                     & Tutorials 2     \\
                     & Tutorials 3     \\    
                     & Tutorials 4     \\
                     & Tutorials 5     \\    
                     & Tutorials 6     \\
                     & Tutorials 7     \\    
                     & Tutorials 8     \\
                     & Tutorials 9     \\    
                     & Tutorials 10    \\
                     & Tutorials 11    \\    
                     & Tutorials 12    \\
                     & Tutorials 13    \\    
                     & Tutorials 14    \\
                     & Tutorials 15    \\    
                     & Tutorials 16    \\
                     & Tutorials 17    \\    
                     & Tutorials 18    \\
                     & Tutorials 19    
\end{tabular}
\end{ruledtabular}
\end{table*}

\begin{figure*}
    \includegraphics[width=\linewidth]{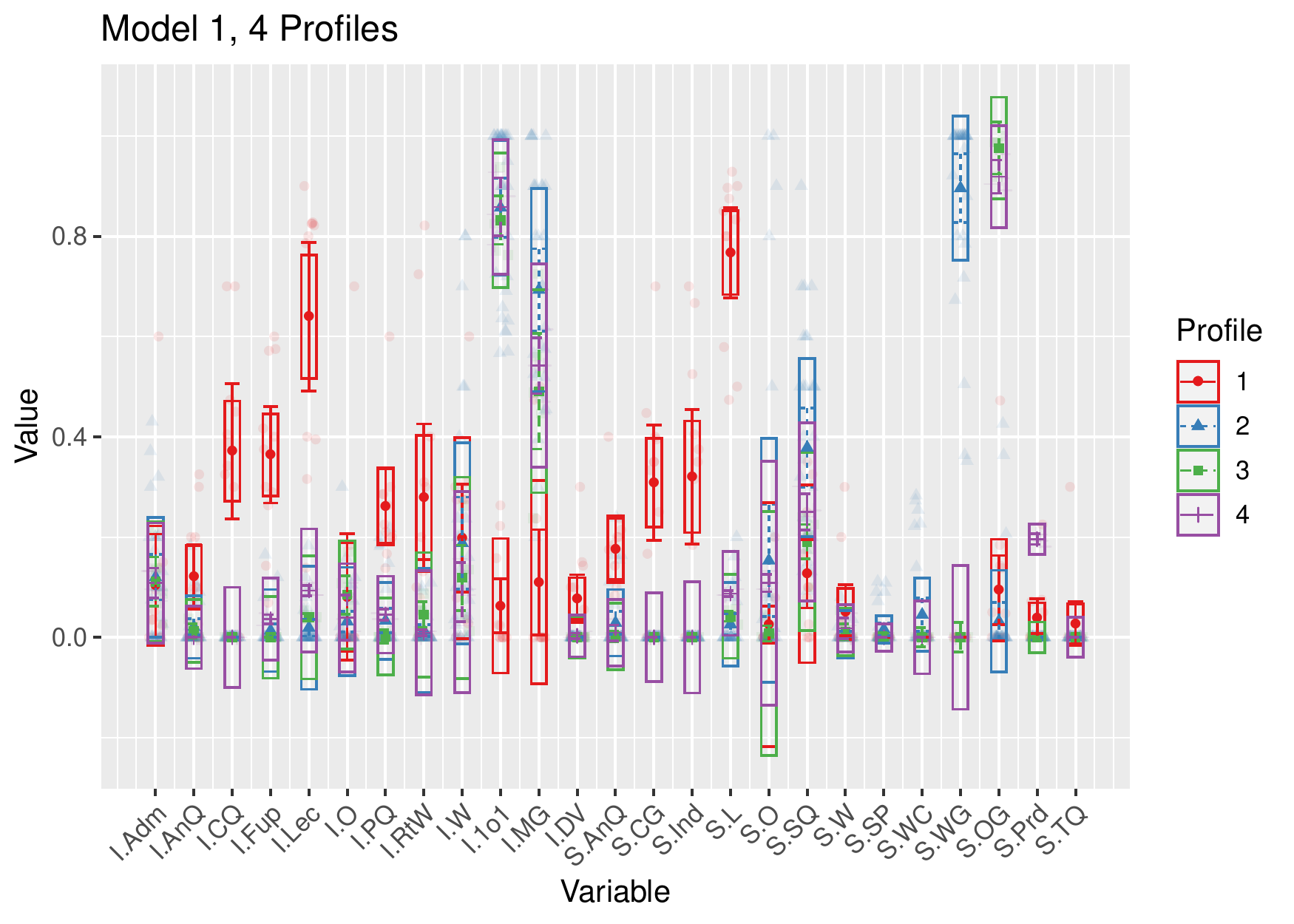}
    \caption{Model 1, 4 profiles. This output kept the sorting from the 3 profile solution, but pulled the ISLE labs into their own category. Profile 3 (in green) holds the ISLE labs, while profile 4 (in purple) holds the context-rich problems labs. The x-axis shows the full set of COPUS codes, while the y-axis shows the fraction of class time dedicated to that code. The light shaded dots show the actual COPUS observations. The boxes show the standard deviation in each assigned profile, and the bars show the confidence interval of the profile centroid of said profile assignment. Since we used the equal variance model, the size of the boxes is equal between classes in each COPUS code.}
    \label{fig:FullList-Model1-4profiles}
\end{figure*}

\FloatBarrier

\begin{table*}
\caption{Model 1, 6 profiles. This output looks similar to the 5 profile solution shown in the main text, but splits the recitation-like observations into two groups. Strangely, one of the Modeling Instruction observations gets pulled into one of the new groups, and three of the tutorial sections are left behind while the rest move into a new group.}
\begin{ruledtabular}
\begin{tabular}{cccccc}
  Class 1    & Class 2  & Class 3 & Class 4  & Class 5 & Class 6  \\
 \hline
 Peer Instruction    &  Modeling 1.1 & Modeling 2.1         & ISLE WC Lab 1 & CRP Lab 1 & Tutorials 1 \\
 SCALE-UP 1.1        &  Modeling 1.2 & ISLE WC Recitation 1 & ISLE LO Lab 1 & CRP Lab 2 & Tutorials 2 \\
 SCALE-UP 1.2        &  Modeling 2.2 & ISLE WC Recitation 2 & ISLE LO Lab 2 & CRP Lab 3 & Tutorials 3 \\
 SCALE-UP 1.3        &  Modeling 3.1 & ISLE WC Recitation 3 & ISLE LO Lab 3 & CRP Lab 4 & Tutorials 4 \\
 ISLE WC Lecture 1.1 &  Modeling 3.2 & ISLE WC Recitation 4 & ISLE LO Lab 4 &           & Tutorials 6 \\
 ISLE WC Lecture 1.2 &               & CRP Discussion 1     & ISLE LO Lab 5 &           & Tutorials 7 \\
 CRP Lecture 1.1     &               & CRP Discussion 2     &               &           & Tutorials 8 \\
 CRP Lecture 1.2     &               & Tutorials 5          &               &           & Tutorials 9 \\
 CRP Lecture 2.1     &               & Tutorials 11         &               &           & Tutorials 10\\
 CRP Lecture 2.2     &               & Tutorials 15         &               &           & Tutorials 12\\
 Tutorials Lecture   &               &                      &               &           & Tutorials 13\\
                     &               &                      &               &           & Tutorials 14\\
                     &               &                      &               &           & Tutorials 16\\
                     &               &                      &               &           & Tutorials 17\\    
                     &               &                      &               &           & Tutorials 18\\
                     &               &                      &               &           & Tutorials 19
\end{tabular}
\end{ruledtabular}
\end{table*}

\begin{figure*}
    \includegraphics[width=\linewidth]{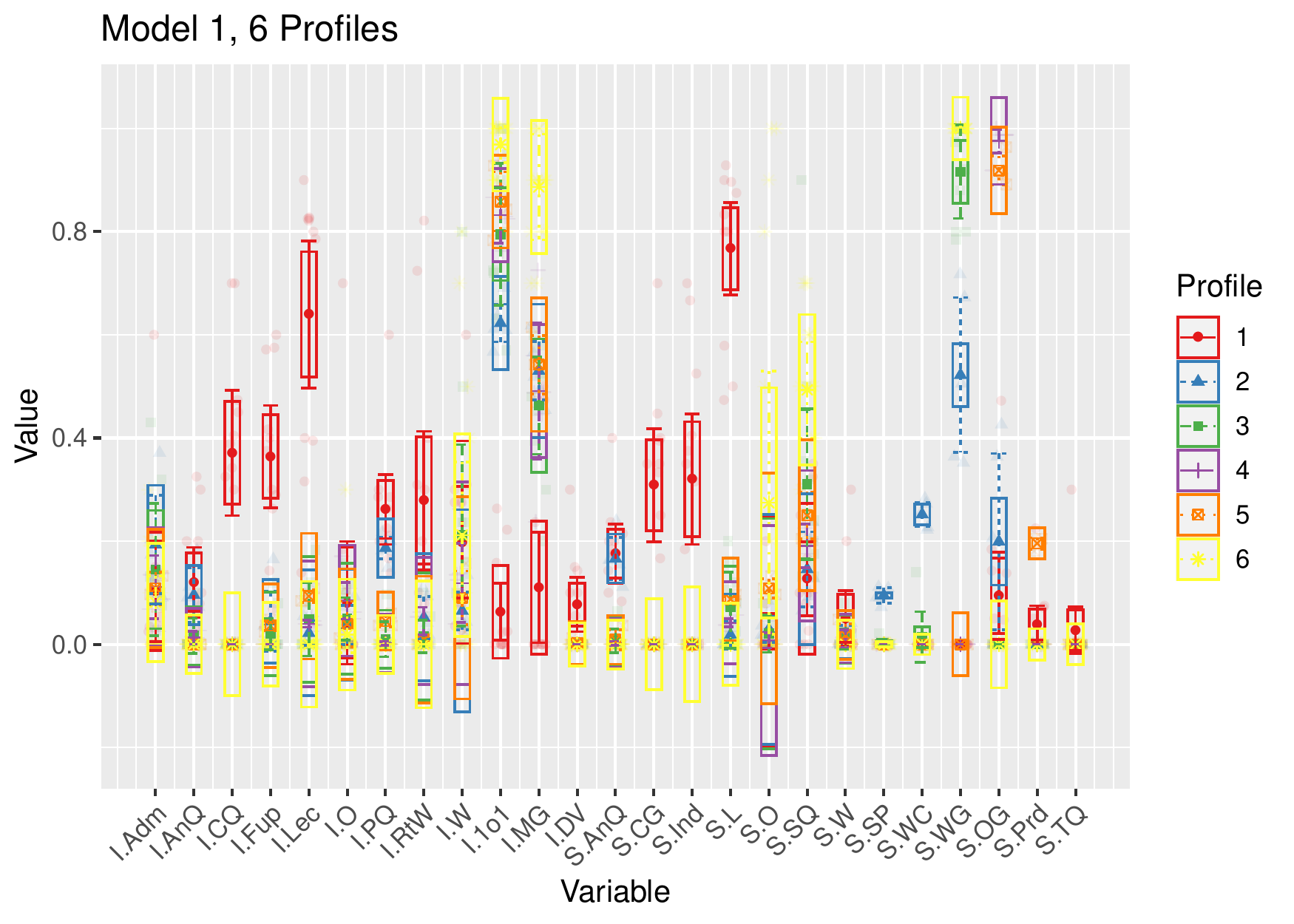}
    \caption{Model 1, 6 profiles. This output looks similar to the 5 profile solution shown in the main text, but splits the recitation-like observations into two groups (profile 3, in green, and profile 6, in yellow). Strangely, one of the Modeling Instruction observations gets pulled into one of the new groups (profile 3), and three of the tutorial sections are left behind while the rest move into a new group (profile 6). The x-axis shows the full set of COPUS codes included in the analysis, while the y-axis shows the fraction of class time dedicated to that code. The light shaded dots show the actual COPUS profiles, colored by their assigned profile. The boxes show the standard deviation in each assigned profile, and the bars show the confidence interval of the centroid of said profile assignment. Since we used the equal variance model, the size of the boxes is equal between classes in each COPUS code.}
    \label{fig:FullList-Model1-6profiles}
\end{figure*}

\FloatBarrier

\begin{table*}
\caption{Model 1, 7 profiles. This output looks similar to the 6 profile solution, but further splits the tutorial observations into two separate groups.}
\begin{ruledtabular}
\begin{tabular}{ccccccc}
  Class 1    & Class 2  & Class 3 & Class 4  & Class 5 & Class 6 & Class 7 \\
 \hline
 Peer Instruction  & Modeling 1.1 & Modeling 2.1         & ISLE WC Lab 1 & CRP Lab 1 & Tutorials 1  & Tutorials 2 \\
 SCALE-UP 1.1      & Modeling 1.2 & ISLE WC Recitation 1 & ISLE LO Lab 1 & CRP Lab 2 & Tutorials 6  & Tutorials 3 \\
 SCALE-UP 1.2      & Modeling 2.2 & ISLE WC Recitation 2 & ISLE LO Lab 2 & CRP Lab 3 & Tutorials 10 & Tutorials 4 \\
 SCALE-UP 1.3      & Modeling 3.1 & ISLE WC Recitation 3 & ISLE LO Lab 3 & CRP Lab 4 & Tutorials 12 & Tutorials 7 \\
 ISLE WC Lecture   & Modeling 3.2 & ISLE WC Recitation 4 & ISLE LO Lab 4 &           & Tutorials 17 & Tutorials 8 \\
 ISLE WC Lecture   &              & CRP Discussion 1     & ISLE LO Lab 5 &           &              & Tutorials 9 \\
 CRP Lecture 1.1   &              & CRP Discussion 2     &               &           &              & Tutorials 13\\
 CRP Lecture 1.2   &              & Tutorials 5          &               &           &              & Tutorials 14\\
 CRP Lecture 2.1   &              & Tutorials 11         &               &           &              & Tutorials 16\\
 CRP Lecture 2.2   &              & Tutorials 15         &               &           &              & Tutorials 18\\
 Tutorials Lecture &              &                      &               &           &              & Tutorials 19
\end{tabular}
\end{ruledtabular}
\end{table*}

\begin{figure*}
    \includegraphics[width=\linewidth]{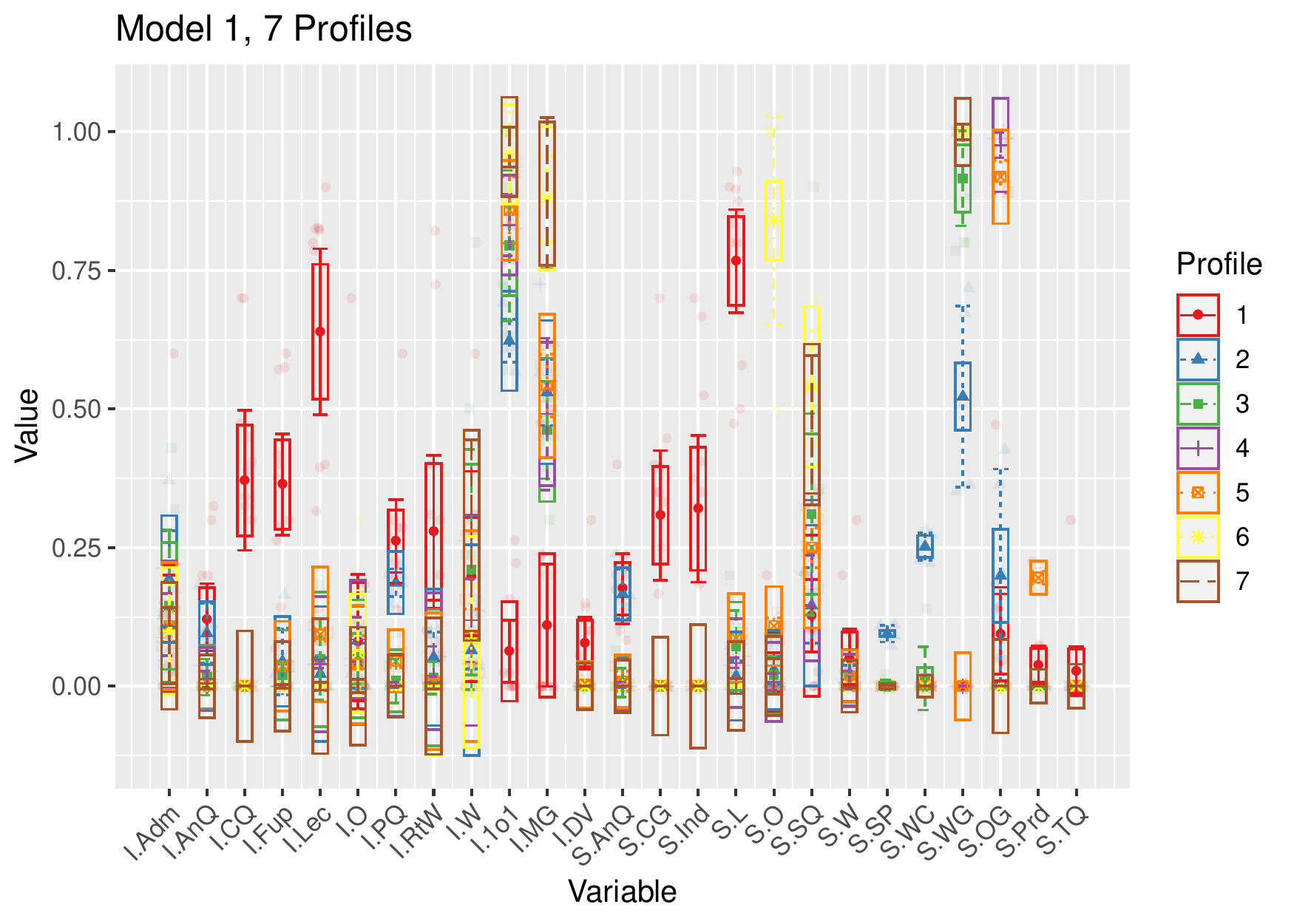}
    \caption{Model 1, 7 profiles. This output looks similar to the 6 profile solution, but further splits the tutorial observations into two separate groups (profile 6, in yellow, and profile 7, in brown). The x-axis shows the full set of COPUS codes included in the analysis, while the y-axis shows the fraction of class time dedicated to that code. The light shaded dots show the actual COPUS profiles, colored by their assigned profile. The boxes show the standard deviation in each assigned profile, and the bars show the confidence interval of the centroid of said profile assignment. Since we used the equal variance model, the size of the boxes is equal between classes in each COPUS code. }
    \label{fig:FullList-Model1-7profiles}
\end{figure*}

\FloatBarrier
\begin{table*}
\caption{Model 1, 8 profiles. The tutorials split even further, and pull in an ISLE recitation. The output is really nonsensical at this point outside of the lecture-like and lab groups.  }
\begin{ruledtabular}
\begin{tabular}{cccccccc}
  Class 1    & Class 2  & Class 3 & Class 4  & Class 5 & Class 6 & Class 7& Class 8 \\
 \hline
 Peer Inst.       & Modeling 1.1 & Modeling 2.1      & ISLE WC Lab 1 & ISLE WC Rec. 4 & CRP Lab 1 & Tutorials 1  & Tutorials 2 \\
 SCALE-UP 1.1     & Modeling 1.2 & ISLE WC Rec. 1    & ISLE LO Lab 1 & Tutorials 3    & CRP Lab 2 & Tutorials 6  & Tutorials 8 \\
 SCALE-UP 1.2     & Modeling 2.2 & ISLE WC Rec. 2    & ISLE LO Lab 2 & Tutorials 4    & CRP Lab 3 & Tutorials 10 & Tutorials 9 \\
 SCALE-UP 1.3     & Modeling 3.1 & ISLE WC Rec. 3    & ISLE LO Lab 3 & Tutorials 7    & CRP Lab 4 & Tutorials 12 & Tutorials 13\\
 ISLE WC Lec. 1.1 & Modeling 3.2 & CRP Disc. 1       & ISLE LO Lab 4 & Tutorials 15   &           & Tutorials 17 & Tutorials 14\\
 ISLE WC Lec. 1.2 &              & CRP Disc. 2       & ISLE LO Lab 5 &                &           &              & Tutorials 16\\
 CRP Lec. 1.1     &              & Tutorials 5       &               &                &           &              & Tutorials 18\\
 CRP Lec. 1.2     &              & Tutorials 11      &               &                &           &              & Tutorials 19 \\
 CRP Lec. 2.1 \\
 CRP Lec. 2.2 \\
 Tutorials Lec. 
\end{tabular}
\end{ruledtabular}
\end{table*}

\begin{figure*}
    \includegraphics[width=\linewidth]{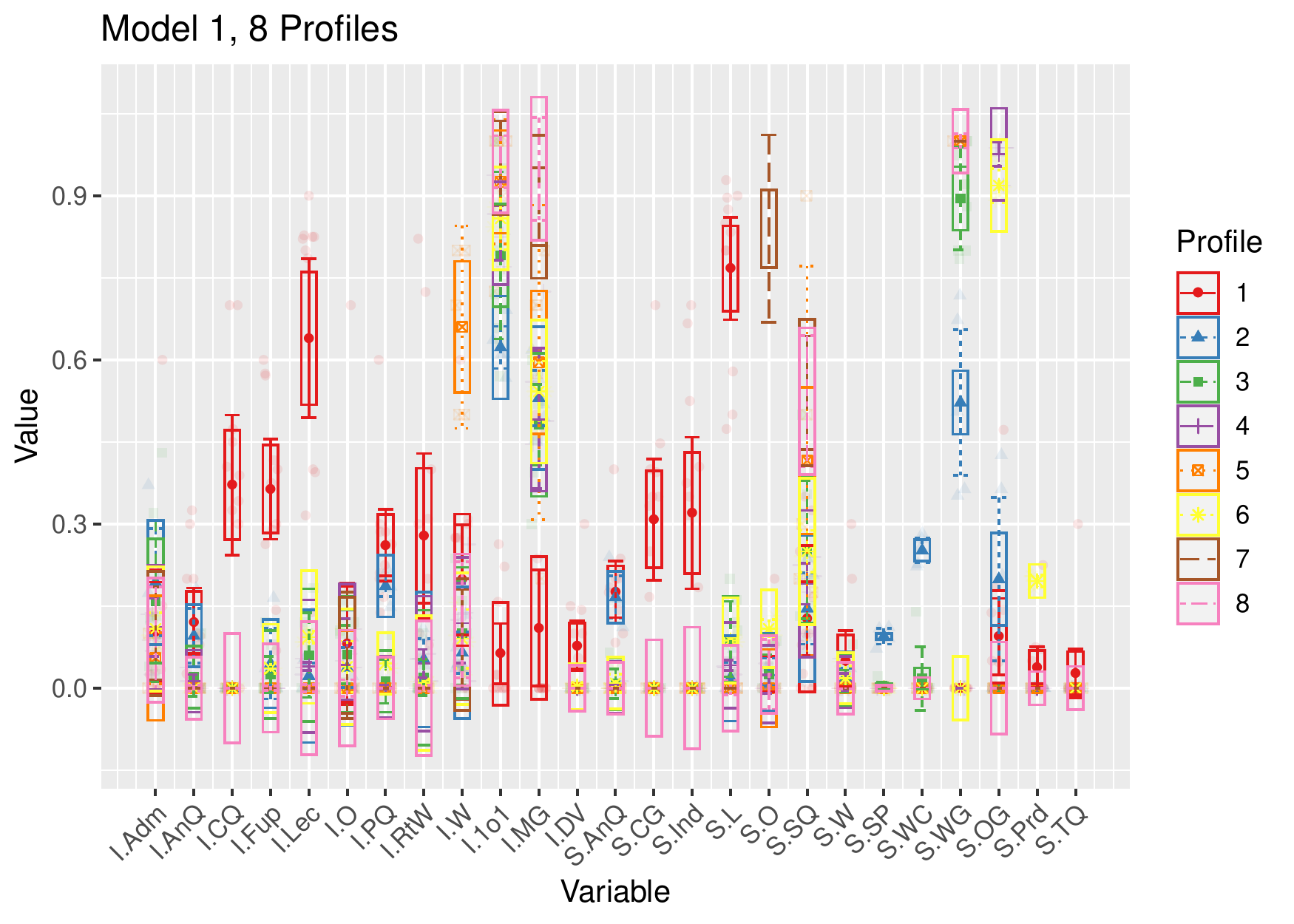}
    \caption{Model 1, 8 profiles. The tutorials split even further, into profiles 5, 7, and 8, and pull an ISLE recitation into profile 5. The output is really nonsensical at this point outside of the lecture-like and lab groups. The x-axis shows the full set of COPUS codes included in the analysis, while the y-axis shows the fraction of class time dedicated to that code. The light shaded dots show the actual COPUS profiles, colored by their assigned profile. The boxes show the standard deviation in each assigned profile, and the bars show the confidence interval of the centroid of said profile assignment. Since we used the equal variance model, the size of the boxes is equal between classes in each COPUS code.}
    \label{fig:FullList-Model1-8profiles}
\end{figure*}

\begin{table*}
  \caption{COPUS Codes for Instructor and Student Activities \cite{Smith2013} \label{tab:COPUSCodes}}
  \begin{ruledtabular}
    \begin{tabular}{ll}
      \textbf{Students are Doing}& \\
      \hline \\
      \textbf{L} & Listening to instructor/taking notes, etc.  \\
      \textbf{Ind} & Individual thinking/problem solving. Only mark when an instructor explicitly asks students to think \\ &about a clicker question or another question/problem on their own  \\
      \textbf{CG} & Discuss clicker question in groups of 2 or more students  \\
      \textbf{WG} & Working in groups on worksheet activity  \\
      \textbf{OG} & Other assigned group activity, such as responding to instructor question  \\
      \textbf{AnQ} & Student answering a question posed by the instructor with rest of class listening  \\
      \textbf{SQ} & Students asks question  \\
      \textbf{WC} & Engaged in whole class discussion by offering explanations, opinions, judgment, etc. to whole class, \\&often facilitated by instructor  \\
      \textbf{Prd} & Making a prediction about the outcome of demo or experiment  \\
      \textbf{SP} & Presentation by student(s)  \\
      \textbf{TQ} & Test or quiz  \\
      \textbf{W} & Waiting (instructor late, working on fixing AV problems, instructor otherwise occupied, etc.)  \\
      \textbf{O} & Other  \\
      
     \hline\\
      \textbf{Instructor is Doing}& \\
      \hline\\
      \textbf{Lec} & Lecturing (presenting content, deriving mathematical results, presenting a problem solution, etc.)  \\
      \textbf{RtW} & Real-time writing on board, document projector, etc. (Often checked off along with Lec)  \\
      \textbf{FUp} & Follow-up/feedback on clicker question or activity to entire class  \\
      \textbf{PQ} & Posing non-clicker question to students (non-rhetorical)  \\
      \textbf{CQ} & Asking a clicker question (mark the entire time the instructor is using a clicker question, not just \\&when first asked)  \\
      \textbf{AnQ} & Listening to and answering student questions with entire class listening  \\
      \textbf{MG} & Moving through class guiding ongoing student work during active learning task \\
      \textbf{1o1} & One-on-one extended discussion with one or more individuals, not paying attention to the rest of \\&class (can be along with MG or AnQ)  \\
      \textbf{D/V} & Showing or conducting a demo, experiment, simulation, video, or animation  \\
      \textbf{Adm} & Administration (assign homework, return tests, etc.)  \\
      \textbf{W} & Waiting when there is an opportunity for an instructor to be interacting with or observing/listening\\& to student or group activities and the instructor is not doing so  \\
      \textbf{O} & Other  \\
    \end{tabular}
  \end{ruledtabular}
\end{table*}